\documentclass[aps,pra,twocolumn,a4paper,amsmath,amssymb,amsthm,showpacs,floatfix,superscriptaddress]{revtex4}

\usepackage{graphicx, color}
\usepackage{epstopdf}
\usepackage{braket}

\newcommand{\Op}[1]{\boldsymbol{\mathsf{\hat{#1}}}}

\newcommand{\Fkt}[1]{\,\mathsf {#1}}

\def\openone{\leavevmode\hbox{\small1\kern-3.3pt\normalsize1}}
\newcommand{\vphi}{\varphi}
\ifx\Tr\renewcommand{\Tr}{\Fkt{Tr}} 
\else\newcommand{\Tr}{\Fkt{Tr}}
\fi

\newtheorem{thm}{Theorem}
\newtheorem{lem}{Lemma}

\begin{document}

\title{Monotonically convergent optimization in quantum control using Krotov's method}
\date{\today}

\author{Daniel M. Reich}
\affiliation{Institut f\"ur Theoretische Physik,
  Freie Universit\"at Berlin,
  Arnimallee 14, 14195 Berlin, Germany}
\affiliation{Theoretische Physik,
  Universit\"at Kassel, Heinrich-Plett-Str. 40, 34132 Kassel, Germany}

\author{Mamadou Ndong}
\affiliation{Institut f\"ur Theoretische Physik,
  Freie Universit\"at Berlin,
  Arnimallee 14, 14195 Berlin, Germany}
\thanks{now at: Laboratoire de Chimie Physique, B\^{a}t. 349, Campus
  d'Orsay, 91405 Orsay Cedex, France}

\author{Christiane P. Koch}
\email{christiane.koch@uni-kassel.de}
\affiliation{Institut f\"ur Theoretische Physik,
  Freie Universit\"at Berlin,
  Arnimallee 14, 14195 Berlin, Germany}
\affiliation{Theoretische Physik,
  Universit\"at Kassel, Heinrich-Plett-Str. 40, 34132 Kassel, Germany}

\date{\today}
\begin{abstract}
  The non-linear optimization method developed by Konnov and
  Krotov [Automation and Remote Control \textbf{60}, 1427 (1999)] has
  been used previously to extend the capabilities of optimal control
  theory from the linear to
  the non-linear Schr\"odinger equation 
  [Sklarz and Tannor, Phys. Rev. A \textbf{66},
  053619 (2002)]. Here we show that based on the Konnov-Krotov method, 
  monotonically convergent algorithms are obtained for a 
  large class of quantum control problems. It includes, 
  in addition to non-linear equations of motion, control
  problems that are characterized by non-unitary time evolution, 
  non-linear dependencies of the Hamiltonian on the control, 
  time-dependent targets  and
  optimization functionals that depend to higher than second order on
  the time-evolving states. 
  We furthermore show that the non-linear (second order) contribution
  can be estimated  either analytically or
  numerically, yielding readily applicable optimization algorithms.
  We demonstrate monotonic convergence for an optimization functional
  that is an eighth-degree polynomial in the states.
  For the 'standard' quantum control problem of a convex
  final-time functional, linear equations of motion and linear
  dependency of the Hamiltonian on the field, the second-order
  contribution is not required for monotonic convergence but can be
  used to speed up convergence.  We demonstrate this by comparing 
  the performance of first and second order
  algorithms for two examples.
\end{abstract}

\maketitle


\section{Introduction}
\label{sec:intro}

Quantum control of light and matter uses
external fields to  engineer constructive and destructive
interferences to steer a physical process into a desired
direction~\cite{RiceBook,BrumerShapiroBook}. The idea was pioneered in
the
1980s~\cite{TannorJCP85,TannorJCP86,BrumerShapiroJCP86,BrumerShapiroCPL86}
and gained wide-spread attention with the advent of femtosecond
lasers and pulse shaping techniques~\cite{WarrenSci93}. It was
realized at about the same time that constructive and destructive
interferences need not be devised by hand but can rather be obtained
employing concepts from engineering such as feedback and
optimization~\cite{Tannor92,JudsonPRL92,SomloiCP93,WarrenSci93,ZhuJCP98}. 
This has significantly
broadened the range of quantum control problems that can be
tackled. In particular, optimal control theory (OCT) has 
become a popular tool employed  in areas as different as
vibrational dynamics of complex molecules~\cite{DoslicPCCP99,WangJCP06},
quantum dots and rings~\cite{HohenesterPRL04,GrossPRL07},
ultracold gases~\cite{SklarzPRA02,MyPRA04,HohenesterPRA07,GrondPRA09},
multi-photon excitations~\cite{MayJCP07,OhtsukiPRA08},
nuclear magnetic resonance~\cite{SkinnerJMR03,KhanejaJMR05} and
quantum information~\cite{JosePRL02,TeschPRL02,TreutleinPRA06}.

All these applications are based on quantum dynamics, but 
differ in their (i) equation of motion, (ii) dependence of the
Hamiltonian on the control, and (iii) target functional. (i) 
In standard applications of OCT to quantum control, the equation of
motion is the linear Schr\"odinger equation, 
e.g., Refs.~\cite{Tannor92,JosePRL02,TeschPRL02,MyPRA04}. 
OCT can also be applied to examples with a non-linear equation of
motion~\cite{SklarzPRA02,WangJCP06,GrondPRA09} which is obtained as an 
effective  description in the framework of the mean-field
approximation.
(ii) The dependence of the Hamiltonian on the control is linear
in most applications of OCT from atomic, molecular and chemical
physics, reflecting a laser field driving optically allowed
transitions. This is changed to a non-linear dependence if
multi-photon excitations~\cite{MayJCP07,OhtsukiPRA08}
or a parametrization of the control field~\cite{GrondPRA09}
is considered. (iii) Linear and bilinear target functionals
have commonly been used in application of OCT to quantum control to
date~\cite{JosePRA03}. However, 
target functionals that are higher order polynomials in the states of
the system may be encountered in quantum information
applications~\cite{Matthias}. 
The type of dynamics and 
functionals translates into different requirements that must be met by
the optimization algorithm. In particular for non-linear equations of
motion, non-linear dependencies of the Hamiltonian on the control,
time-dependent targets where the target operator is non-semidefinite and
target functionals that are higher order polynomials in the states, 
it is not straightforward to construct \textit{monotonically}
convergent algorithms. This is the problem that we address here. 

We utilize the non-linear optimization algorithm by Konnov and
Krotov~\cite{Konnov99} which had been translated to the language of
quantum mechanics and first employed for solving a quantum control
problem by Sklarz and Tannor~\cite{SklarzPRA02}. Specifically, 
Sklarz and Tannor realized that a generalized form of the
optimization functional yielding modified adjoint states is
required in order to apply OCT
to a non-linear Schr\"odinger equation~\cite{SklarzPRA02}.  
We show that the work by Sklarz and Tannor applies also to quantum
control problems that are characterized by non-unitary time evolution,
time-dependent targets, 
non-linear dependencies of the Hamiltonian on the control
and target functionals that depend to higher than second order on the
time-evolving states. Translating the original proof by Konnov and
Krotov~\cite{Konnov99} to quantum control, we furthermore show 
that the parameters of the optimization algorithm can be
estimated analytically or numerically. 
We can thus give a ready-to-use prescription of the algorithm 
where the parameters are determined by the physics of the problem. 
For the case of a quadratic functional and linear equations of
motion, a first order construction of Krotov's method is sufficient to
guarantee monotonic
convergence~\cite{SklarzPRA02,JosePRA03}. We show that in this case, a
second order contribution may still be used to speed up convergence.   

The paper is organized as follows.
The optimization algorithm and the estimate of the algorithm's
parameters are presented in Section~\ref{sec:OCT} with all 
mathematical details of the derivation found in the appendices.
Sections~\ref{sec:J_LI} and~\ref{sec:appl} are devoted to 
applications of the optimization algorithm. We demonstrate monotonic
convergence for one example that requires a second order construction
in Section~\ref{sec:J_LI}, and we present two examples in
Section~\ref{sec:appl} for which the second order construction is not
required but may be used to speed up convergence. 
Section~\ref{sec:concl} concludes.


\section{Optimization algorithm}
\label{sec:OCT}

\subsection{Control problem}
\label{subsec:problem}

The control problem is characterized by stating the control target and
possible additional costs in functional form,
\begin{eqnarray}
  \label{eq:J}
  J =
  J_T\left[\{\vphi_k(T)\}\right]  
  + \int_0^T J_t\left[\{\vphi_k(t)\},\epsilon(t)\right]\;dt \,,
\end{eqnarray}
where we have separated final-time $T$ and intermediate-time $t$ 'costs'.
$\{\vphi_k(t)\}$
denotes a set of complex state vectors~\footnote{
  In principle, all functionals depend on the state vectors
  and their complex conjugates, $|\vphi_k(t)\rangle$ and 
  $\langle\vphi_k(t)|$, and   differentiation needs to be carried out
  with respect to the state 
  vectors and their complex conjugates.
  To simplify notation, we indicate by $\varphi_k(t)$ the dependence
  on both the bra and the ket vectors.
  }
and $\epsilon(t)$ is the external field or control that shall be optimized.
The final-time cost is typically specified in terms of
some desired unitary operator $\Op{O}$, for example~\cite{JosePRA03},
\begin{widetext}
\begin{align}
  \label{eq:J_T}
    J_T\left[\{\vphi_k(T)\}\right] &=
    -\frac{\lambda_0}{N^2}\left|\Tr\left\{
      \Op{O}^{\dagger}\Op{P}_N\Op{U}(T,0;\epsilon)\Op{P}_N\right\}\right|^2 \\
   &= \nonumber
  -\frac{\lambda_0}{N^2}
    \sum_{k,k'=1}^N \langle \vphi_k(t=0)|
    \Op{O}^{\dagger}\Op{U}(T,0;\epsilon)|\vphi_k(t=0)\rangle
    \langle \vphi_{k'}(t=0)|
    \Op{U}(T,0;\epsilon)^{\dagger}\Op{O}|\vphi_{k'}(t=0)\rangle
  \,.
\end{align}
\end{widetext}
Here, $\Op{U}(T,0;\epsilon)$ denotes the time evolution operator from
the initial time $t=0$ to the final time $T$ under
the action of the field $\epsilon$, and $\lambda_0$ is a weight.
The dimension of the subspace of the total Hilbert space,
$\mathcal{H}$, on which the target operator $\Op{O}$ acts, $\mathcal{H}_O$,
is denoted by $N$, and $\Op{P}_N$
is the projector onto $\mathcal{H}_O$. $\{|\vphi_k(t=0)\rangle\}\equiv\{|k\rangle\}$
is a suitable orthonormal basis spanning this subspace.
A single state-to-state transition is  obtained by taking $\Op{O}$ to
be the projector onto the target state, i.e. $N=1$. In order to optimize
one-qubit or two-qubit quantum gates, $N=2$ and $N=4$, respectively.
The functional is normalized by $1/N^2$ such that the optimum
corresponds to the weight $J_T=-\lambda_0$. 

The functional of Eq.~(\ref{eq:J_T}) is quadratic in the states at
final time $\{\vphi_k(T)\}$. A linear dependence is obtained
by taking the real part instead of the square modulus of the
trace and yields a phase-sensitive functional~\cite{JosePRA03}. 
Generally, expressing the functionals in terms of expectation values
yields at most a quadratic dependence on the states.
A functional that is a higher order polynomial in the states is
obtained in the context of quantum information when
optimizing for a certain degree of entanglement rather
than a specific unitary transformation~\cite{Matthias}, see 
Section~\ref{sec:J_LI} below.

The intermediate-time cost $J_t$,
\begin{eqnarray}
  \label{eq:J_t}
  J_t\left[\{\vphi_k(t)\},\epsilon\right] &=& 
  g\left[\{|\vphi_k(t)\rangle\},\epsilon(t),t\right]
  \nonumber \\ &=&
  g_a\left[\epsilon(t),t\right] +
g_b\left[\{\vphi_k(t)\},t\right]
\,,
\end{eqnarray}
 is typically used to minimize the
field intensity and to switch the field smoothly on and off,
\begin{equation}
  \label{eq:g_a}
  g_a\left[\epsilon(t)\right] =
  \frac{\lambda_a}{S(t)} \left[ \epsilon(t) -
    \epsilon_\mathrm{ref}(t)\right]^2\,,
\end{equation}
where $\epsilon_\mathrm{ref}(t)$ denotes some reference field, $S(t)$
is a positive (shape) function  and $\lambda_a$ a
weight. $J_t$ can
also be used to formulate time-dependent
targets \cite{OhtsukiJCP04,KaiserJCP04,SerbanPRA05} or
constraints that depend on the state of the
system at intermediate times \cite{JoseMyPRA08} such as
\begin{equation}
  \label{eq:g_b}
  g_b\left[\{\vphi_k(t)\}\right] =
  \frac{\lambda_b}{T N} \sum_{k=1}^N
  \langle \vphi_k(t)|\Op{D}(t)|\vphi_k(t)\rangle\,,
\end{equation}
where the dependence on the states again is quadratic.
While complicated dependencies of $g_a\left[\epsilon\right]$ and
$g_b\left[\{\vphi_k\}\right]$ are conceivable,
we require  $g_a\left[\epsilon\right]$ and
$g_b\left[\{\vphi_k\}\right]$  to be additive,
cf. Eq.~\eqref{eq:J_t}. 
This assumption is typically justified 
since costs or penalties involving the field are usually not related
to costs concerning the dynamics of the system.

The time evolution operator required to evaluate the functional, 
Eq.~(\ref{eq:J_T}), can be obtained by solving the equation of motion
for each of the basis states, 
\begin{eqnarray}
  \label{eq:eom}
  \frac{d}{dt} |\vphi_k(t)\rangle &=& -\frac{i}{\hbar}\Op{H}[\vphi_k,\epsilon]
  |\vphi_k(t)\rangle \;, \nonumber \\
  &=& |f_k(\vphi_k,\epsilon)\rangle \;,\;k=1,\ldots,N\,.
\end{eqnarray}
An explicit dependency of the Hamiltonian on the state,
$\Op{H}[\vphi_k]$, will occur for non-linear Schr\"odinger
equations such as the Gross-Pitaevski equation or Hartree-Fock-like
equations where the Hamiltonian is second order in the
states~\cite{SklarzPRA02,WangJCP06,GrondPRA09,MundtNJP09}.
The dependency of the Hamiltonian on the field can be linear,
corresponding to one-photon dipole coupling, or higher order for
non-resonant multi-photon transitions. 
Equation~\eqref{eq:eom} can be extended to account for
non-unitary time evolution by considering the density operator to be a
vector in Liouville space and replacing the Hamiltonian by the
Liouvillian~\cite{MukamelBook}.

\subsection{Optimization equations}

The essence of Krotov's method~\cite{Konnov99,SklarzPRA02} 
consists in disentangling the interdependency of
the states and the control by 
constructing an auxiliary
functional $L[\{\vphi_k\},\epsilon,\Phi]$ that depends on the states,
the control and an arbitrary scalar potential $\Phi$ such that
for any $\Phi$, $L[\{\vphi_k\},\epsilon,\Phi]$ and $J[\{\vphi_k\},\epsilon]$
are identical and minimization of $L[\{\vphi_k\},\epsilon,\Phi]$ 
is completely equivalent to minimization of $J[\{\vphi_k\},\epsilon]$.
This is achieved by adding a vanishing quantity,
\begin{eqnarray}
  L[\{\vphi_k\},\epsilon,\Phi] &=& \nonumber
  G(\{\vphi_k(T)\}) 
  -\Phi(\{\vphi_k(0)\},0)  \\
  &&- \int_0^T  R(\{\vphi_k(t)\},\epsilon(t),t) dt 
  \label{eq:defL}  
\end{eqnarray}
with 
\begin{eqnarray*}
  \label{eq:defG}
  G\left[\{\vphi_k(T)\}\right]  &=&
  J_T[\{\vphi_k(T)\}] 
  + \Phi(\{\vphi_k(T)\},T)\,,\\
  R\left[\{\vphi_k(t)\},\epsilon(t),t\right] &=&
  -\left(g_a[\epsilon(t),t]+g_b[\{\vphi_k(t)\},t]\right)
    \nonumber \\ &&
  + \frac{\partial \Phi}{\partial t} + \sum_{k=1}^{N} 
  \nabla_{|\vphi_k\rangle} \Phi \cdot
  |f_k[\vphi_k,\epsilon,t]\rangle \\ &&
  + \sum_{k=1}^{N}
  \langle f_k[\vphi_k,\epsilon,t]| \cdot
  \nabla_{\langle\vphi_k|} \Phi\,.
  \label{eq:defR}
\end{eqnarray*}
It is sufficient to expand the functional
$\Phi(\{|\vphi_k\rangle\},t)$  up to second order in the states to ensure
the necessary monotonicity conditions for arbitrary change of the
state vectors, 
\begin{eqnarray}
  \label{eq:Phinew}
  \Phi(\{\vphi_k\},t) &=&  \sum_k \bigg[
  \langle\chi_k(t)|\vphi_k(t)\rangle + 
  \langle\vphi_k(t)|\chi_k(t)\rangle\bigg] \nonumber \\
  &&+\frac{1}{2}\sum_{kl}\langle
  \Delta\vphi_k(t)|\Op{\sigma}(t)|\Delta\vphi_l(t)\rangle\,, 
\end{eqnarray}
with first order expansion coefficients $\chi_k(t)$ and
$\Delta\vphi_k(t)$ the change in the
state $\vphi_k(t)$, given by 
$\Delta\vphi_k(t)=\vphi_k^{(i+1)}(t)-\vphi_k^{(i)}(t)$ 
where the superscripts $(i)$ and $(i+1)$
denote propagation under the old and new fields, $\epsilon^{(i)}(t)$ and
$\epsilon^{(i+1)}(t)$, respectively. Due to the freedom of choice in
$\Phi$, the operator
$\Op\sigma(t)$ in the second order term can be chosen to ensure the 
extremum condition with respect to the state changes.
If $J$ is to be minimized, $\Op\sigma(t)$ will be chosen to
\textit{maximize} it with respect to the change in the states. Any change in
the field leads then to monotonic decrease of $J$~\cite{Konnov99,SklarzPRA02}. 

Requiring the first order derivatives of $L$ with respect to 
$|\vphi_k(t)\rangle$, $\langle\vphi_k(t)|$ and $\epsilon(t)$
to vanish, yields a set of coupled control equations for 
the first order expansion coefficients
$\chi_k(t)$ and the field. 
Since all involved functionals are real, it
is sufficient to state the equation for the kets, the bra equations
being given by the adjoints,
\begin{widetext}
\begin{subequations}\label{eq:backward_prop}
  \begin{eqnarray}\label{eq:chidot}
  \frac{d}{dt}|\chi^{(i)}_k(t)\rangle &=&
  -\frac{i}{\hbar}\Op{H}^{\dagger}[\varphi^{(i)}_k,\epsilon^{(i)}]
  |\chi^{(i)}_k\rangle 
  -\frac{i}{\hbar} \left[
    \sum_l \big\langle \vphi_l^{(i)} \big| \nabla_{\Ket{\vphi_k}} 
    \Op{H}^\dagger[\varphi^{(i)}_k,\epsilon^{(i)}]
    \big|\chi_l^{(i)}\big\rangle -
    \sum_l \big\langle \chi_l^{(i)} \big| \nabla_{\Bra{\vphi_k}} 
    \Op{H}[\varphi^{(i)}_k,\epsilon^{(i)}]
    \big|\vphi_l^{(i)}\big\rangle 
  \right] \nonumber \\ &&
  + \nabla_{\Bra{\vphi_k}} g_b\big|_{|\vphi^{(i)}_k(t)\rangle} \,,\\
  |\chi^{(i)}_k(T)\rangle &=& -\nabla_{\Bra{\vphi_k}}
  J_T\big|_{|\vphi^{(i)}_k(T)\rangle} 
  \,,\quad k=1,\ldots, N \,. \label{eq:chiT}
  \end{eqnarray}
\end{subequations}
\end{widetext}
The 'initial' condition at the final time $T$ is given in terms of the
gradient with respect to the states of the final-time cost,
$J_T$. Note that all gradients in Eqs.~\eqref{eq:backward_prop}
are evaluated with the states
$\varphi_k^{(i)}$ that are forward propagated under the old field,
$\epsilon^{(i)}$ as indicated by the superscript $(i)$. 
The prescription for the new field is obtained by evaluating the
derivative of the constraints with respect to the field, 
\begin{widetext}
  \begin{equation}\label{eq:newfield}
    \frac{\partial g}{\partial
      \epsilon}\bigg|_{\epsilon^{(i+1)},|\vphi^{(i+1)}\rangle}
    = 2\mathfrak{Im} \left[\sum_{k=1}^N \left\langle \chi_k^{(i)}(t)\Bigg|
      \frac{\partial \Op{H}}{\partial \epsilon}
      \bigg|_{\epsilon^{(i+1)},\vphi^{(i+1)}}
      \Bigg| \vphi_k^{(i+1)}(t)\right\rangle +
    \frac{1}{2}\sigma(t)\sum_{k=1}^N
    \left\langle \Delta\vphi_k(t) \Bigg|
      \frac{\partial\Op{H}}{\partial\epsilon}\bigg|_{\epsilon^{(i+1)},\vphi^{(i+1)}}
      \Bigg|\vphi_k^{(i+1)}(t)\right\rangle \right] \,.
  \end{equation}
\end{widetext}
It involves backward propagation of the
adjoint states under the old field, $\chi_k^{(i)}(t)$, 
and forward propagation of the states, $\varphi_k^{(i+1)}(t)$, under
the new field,
\begin{subequations}\label{eq:forward_prop}
\begin{eqnarray}
  \frac{d}{dt}|\vphi^{(i+1)}_k(t)\rangle &=&
  -\frac{i}{\hbar}\Op{H}[\vphi_k^{(i+1)},\epsilon^{(i+1)}]
  |\vphi^{(i+1)}_k(t)\rangle \\
  |\vphi^{(i+1)}_k(0)\rangle &=& |k\rangle \,,\quad k=1,\ldots, N   \,.
\end{eqnarray}
\end{subequations}
Equations~\eqref{eq:backward_prop}-\eqref{eq:forward_prop} need to be solved
simultaneously.
The iteration is started by propagating  Eqs.~(\ref{eq:forward_prop})
under some guess field, $\epsilon^{(0)}(t)$, to obtain
$\vphi^{(0)}_k(T)$  and evaluate  Eq.~\eqref{eq:chiT}.
The equation for the backward propagation, Eq.~\eqref{eq:chidot}, 
becomes an inhomogeneous Schr\"odinger equation for non-linear
equations of motion, cf. the terms in square brackets, or if the 
intermediate-time cost, $g_b$, depends on the states
($\nabla_{\Bra{\vphi_k}} g_b \neq 0$). If
the time-dependent cost over the field takes the form
of Eq.~(\ref{eq:g_a}), the equation for the new field reads
\begin{widetext}
  \begin{eqnarray}\label{eq:neweps}
    \epsilon^{(i+1)}(t) &=&
    \epsilon_\mathrm{ref}(t) + \\ && \nonumber
    \frac{S(t)}{\lambda_a}
    \mathfrak{Im} \left\{\sum_{k=1}^N \left\langle \chi_k^{(i)}(t)\Bigg|
      \frac{\partial \Op{H}}{\partial \epsilon}
      \bigg|_{\epsilon^{(i+1)},\vphi^{(i+1)}}
      \Bigg| \vphi_k^{(i+1)}(t)\right\rangle +
  \frac{1}{2}  \sigma(t)\sum_{k=1}^N
    \left\langle \Delta\vphi_k(t) \Bigg|
      \frac{\partial\Op{H}}{\partial\epsilon}\bigg|_{\epsilon^{(i+1)},\vphi^{(i+1)}}
      \Bigg|\vphi_k^{(i+1)}(t)\right\rangle \right\} \,.
  \end{eqnarray}
\end{widetext}
Moreover, for dipole transitions the Hamiltonian is given by
$\Op{H}=\Op{H}_0+\Op{\mu}\epsilon(t)$;
and hence $\partial \Op{H}/\partial \epsilon=\Op{\mu}$. 
We thus recover
the familiar prescription for the change in field obtained for a first
order construction of the algorithm~\cite{JosePRA03}
plus an additional second order contribution,  given in terms of
the change in the states,
$\Delta\vphi_k^{(i+1)}(t)$, with 'weight' $\sigma(t)$.

A choice of $\sigma(t)$ that 
guarantees a maximum of $L$ with respect to the states (i.e., a
positive second derivative of $R$ and negative second derivative of $G$)
is given by~\cite{Konnov99}
\begin{subequations}  \label{eq:sigma}
\begin{eqnarray}
  \sigma(t) &=&
  e^{\bar{B}(T-t)}\left(\frac{\bar{C}}{\bar{B}}-\bar{A}\right) -
  \frac{\bar{C}}{\bar{B}} \quad \mathrm{for} \quad \bar{B} \neq 0\,, \\
  \sigma(t) &=& \bar{C}(T-t)-\bar{A} \quad\quad\quad\quad\quad
  \mathrm{for} \quad \bar{B} = 0\,.
\end{eqnarray}
\end{subequations}
The physics of the problem, i.e., the dependency of the functional on
the states, the dependency of the Hamiltonian on the control,
(non-)linearity of the  equation of motion governing and unitary or
non-unitary of the time evolution, determine the 
parameters $\bar A$, $\bar B$ and $\bar C$, 
\begin{subequations}  \label{eq:bar}
\begin{eqnarray}
  \bar{A} &=&\max\left(\varepsilon_A,2A+\varepsilon_A\right)\,,\\
  \bar{B} &=& 2B+\varepsilon_B\,,\\
  \bar{C} &=&\min\left(-\varepsilon_C,2C-\varepsilon_C\right)\,,
\end{eqnarray}
\end{subequations}
where the $\varepsilon_i$ are non-negative numbers that can be used to
enforce strict inequality~\footnote{ 
Note that the choice $\varepsilon_i=0$ ($i=A,B,C$), 
in the worst case, allows 
for iterations in which no change towards the objective due to
the change in the state variables is achieved. However, we will
ensure strict monotonic convergence with respect to the change in the
field such that in practice even the choice $\varepsilon_i=0$ will
almost always lead to strict monotonic convergence.}. 
It should be emphasized, that only in cases where $A$, $B$ and $C$ all
turn out to be zero, the linear 
version of Krotov's method~\cite{JosePRA03} is sufficient to guarantee
monotonic convergence. 

\subsection{Parameters of the second order contribution}
\label{subsec:ABC}

For quantum control problems
the parameters $A$, $B$ and $C$ can either be estimated analytically
or approximated numerically by considering 
$\Delta G = G(\{\vphi_k^{(i+1)}(T)\})
-G(\{\vphi_k^{(i)}(T)\})$
and $\Delta R = R(\{\vphi_k^{(i+1)}(t)\},\epsilon^{(i)}(t),t)
-R(\{\vphi_k^{(i)}(t)\},\epsilon^{(i)}(t),t)$ and guaranteeing
their negativity and positivity, respectively. 
The analytical estimate is based on a
worst case scenario and strictly guarantees monotonic convergence. The
worst case scenario may, however, not always be required and a more
efficient while less fail-proof approach is given by numerical
approximation of $A$, $B$ and $C$.

The analytical estimate of $A$ is obtained by requiring 
$\Delta G \le 0$. This is guaranteed if
\begin{widetext}
\begin{eqnarray}
  \label{eq:defA}
  A=\sup_{\{\Delta\vphi_k\}}
  \frac{\sum_{k=1}^{N}\left[\Braket{\chi_k(T)|\Delta\vphi_k(T)}+
      \Braket{\Delta\vphi_k(T)|\chi_k(T)}\right]
    +J_T\left(\{\vphi_k^{(i)}(T)+\Delta\vphi_k(T)\}\right)
    -J_T\left(\{\vphi_k^{(i)}(T)\}\right)}
  {\sum_{k=1}^{N}\left[\Braket{\Delta\vphi_k(T)|\Delta\vphi_k(T)}\right]}  \,,
\end{eqnarray}
\end{widetext}
where the supremum
needs to be taken over all sets of possible state change vectors
$\{\Delta\vphi_k(T)\}$ with norm 
$\sum_{k=1}^{N}\Braket{\Delta\vphi_k(T)|\Delta\vphi_k(T)}$
between zero and $2N$.
Konnov and Krotov proved~\cite{Konnov99} that, under certain conditions which are almost
trivially fulfilled for quantum systems, quantities like the one on the right-hand
side of Eq.~\eqref{eq:defA} and similar ones obtained for $B$ and $C$ exist
and are well-defined, see also Appendix~\ref{sec:proof}. We discuss in
Appendix~\ref{sec:adaptproof} how 
Konnov's and Krotov's proof simplifies for quantum systems such that
the supremum in Eq.~\eqref{eq:defA} can be estimated. Specifically, 
we show in Appendix~\ref{sec:ABC} that the argument of the supremum in
Eq.~\eqref{eq:defA}  can be rewritten in terms
of a Taylor series of $J_T$ that starts at the second
order. The series can be estimated by its Lagrange remainder,
\begin{equation}
  \label{eq:A_est}
  A \le \frac{1}{2} \sup_{\{\Delta\vphi_k\};\left|\alpha\right|=2 }
  \partial^{\alpha} J_T({\{\Delta \vphi_k(T)}\}) \,,
\end{equation}
i.e., $A$ is given by the supremum over the second derivatives of the
final-time functional, $J_T$,
with respect to the states, $\vphi_k(T)$. The multi-index $\alpha$ and
derivative $\partial^\alpha$ are defined in Eqs.~\eqref{eq:alpha} and
\eqref{eq:dalpha}, respectively. 
For functionals $J_T$ that are linear
or convex in $\vphi_k(T)$, i.e. those for which the second derivatives vanish
or are always non-positive, $A \le 0$. For simplicity one can then
choose $A = 0$, $\varepsilon_A=0$ such that $\bar{A}=0$.

The analytical estimates of $B$ and $C$ are obtained  by requiring $\Delta R
\ge 0$ where
\begin{widetext}
\begin{eqnarray}
  \label{eq:DeltaRnonzero}
  \Delta R\left(\{\Delta\vphi(t)\},t\right) &=&
  \sum_{k=1}^{N}\left[\Braket{\Delta\vphi_k(t)|\Delta\vphi_k(t)}\right]\Bigg[
  \frac{1}{2}\dot{\sigma}(t) + \sigma(t)
  \frac{\sum_{k=1}^{N}
    \left[\Braket{\Delta\vphi_k(t)|\Delta f_k(t)}+
      \Braket{\Delta f_k(t)|\Delta\vphi_k(t)}\right]}
  {\sum_{k=1}^{N}\left[\Braket{\Delta\vphi_k(t)|\Delta\vphi_k(t)}\right]}\nonumber\\
  &&+\frac{\sum_{k=1}^{N}\left[\Braket{\dot{\chi}_k(t)|\Delta\vphi_k(t)}
      +\Braket{\Delta\vphi_k(t)|\dot{\chi}_k(t)}+
      \Braket{\chi_k(t)|\Delta f_k(t)}+\Braket{\Delta f_k(t)|\chi_k(t)}\right]-\Delta g}
  {\sum_{k=1}^{N}\left[\Braket{\Delta\vphi_k(t)|\Delta\vphi_k(t)}\right]}
\Bigg]\,.
\end{eqnarray}
\end{widetext}
The second summand in the square brackets determines $B$ and the third
one $C$. 
$\Delta f_k$ and $\Delta g$
describe the change, due to changes in the states,
in the equations of motion, 
\begin{eqnarray}
  |\Delta f_k(\Delta \vphi_k,t)\rangle & = &
  |f_k(\vphi_k^{(i)}(t)+\Delta\vphi_k(t),\epsilon^{(i)}(t),t\rangle
  \nonumber \\ &&
  -|f_k(\vphi_k^{(i)}(t),\epsilon^{(i)}(t),t)\rangle \,,\label{eq:deltaf}
\end{eqnarray}
and in the constraint,
\begin{eqnarray}
  \Delta g(\{\Delta\vphi_k\},t) & = &
  g(\{\vphi_k^{(i)}(t)+\Delta\vphi_k(t)\},\epsilon^{(i)}(t),t)\nonumber
  \\ &&
  -g(\{\vphi_k^{(i)}(t)\},\epsilon^{(i)}(t),t)\,.\label{eq:deltag}
\end{eqnarray}
$B$ is given by 
\begin{widetext}
\begin{eqnarray}
  B &=& \sup_{\{\Delta\vphi_k\};t\in\left[0,T\right]}
  \left|\frac{\sum_{k=1}^{N}\left[\Braket{\Delta\vphi_k(t)|\Delta f_k(t)}+\Braket{\Delta f_k(t)|\Delta\vphi_k(t)}\right]}{\sum_{k=1}^{N}\left[\Braket{\Delta\vphi_k(t)|\Delta\vphi_k(t)}\right]}\right|
  \label{eq:defB} 
\end{eqnarray}
\end{widetext}
and can be rewritten in terms of a supremum over the Taylor expansion of the
Hamiltonian starting at first order and a supremum over the
Hamiltonian, cf. Appendix~\ref{sec:ABC}. Estimating the Taylor series by its
Lagrange remainder, we obtain 
\begin{widetext}
\begin{eqnarray}
  B &\le& 2\sqrt{N}
    \sup_{\begin{subarray}{c} \Delta\vphi_k;|\alpha|=1 \\ t\in[0,T] \end{subarray}}\left|
      \partial^{\alpha} \Op H\left(\Delta\vphi_k(t)\right)\right| +
  2 \sup_{\{\Delta\vphi_k\};t\in\left[0,T\right]}
  \left|\frac{\sum_{k=1}^N \mathfrak{Im} 
    \Braket{\Delta\vphi_k(t)|\Op H\left(\Delta\vphi_k(t),\epsilon^{(i)},t\right)|\Delta\vphi_k(t)}}
    {\sum_{k=1}^{N}\left[\Braket{\Delta\vphi_k(t)|\Delta\vphi_k(t)}\right]}\right|,
  \label{eq:Bfinal}
\end{eqnarray}
\end{widetext}
i.e., $B$ can be estimated by the supremum over the first derivative of the
Hamiltonian with respect to the states $\vphi(t)$ and a term which can be interpreted as
twice the maximum absolute value of the imaginary part of the Hamiltonian's eigenvalues.
For unitary time evolution governed by the standard
(linear) Schr\"odinger equation, $B=0$ can easily be proven 
and in this case $\sigma(t)$ is a linear function of
time. It is also possible to take $B=0$ for non-unitary time evolution
with linear equations of motion provided $A$ and $C$ can taken 
to be zero. For non-linear equations of motion, the supremum of the first
order derivatives of the Hamiltonian needs to be evaluated explicitly.
$C$ is found to be 
\begin{widetext}
\begin{eqnarray}
    C &=& \inf_{\{\Delta\vphi_k\};t\in\left[0,T\right]}
    \frac{\sum_{k=1}^{N}\left[\Braket{\dot{\chi}_k(t)|\Delta\vphi_k(t)}+\Braket{\Delta\vphi_k(t)|\dot{\chi}_k(t)}
          +\Braket{\chi_k(t)|\Delta f_k(t)}+\Braket{\Delta f_k(t)|\chi_k(t)}\right]-\Delta
      g}
      {\sum_{k=1}^{N}\left[\Braket{\Delta\vphi_k(t)|\Delta\vphi_k(t)}\right]} \label{eq:defC}\,.
\end{eqnarray}
As shown in Appendix~\ref{sec:ABC}, it can be rewritten in terms of suprema
of the Taylor series of the Hamiltonian and  the constraint, $g$,
starting at first and second order, respectively. Estimating the
Taylor series by their Lagrange remainder, we obtain 
\begin{eqnarray}
  \label{eq:Cfinal}
   -C &\ge& 2\sum_{k=1}^{N}\left[\sqrt{\Braket{\chi_k^{(i)}(t)|\chi_k^{(i)}(t)}} \cdot \sup_{\begin{subarray}{c} \Delta\vphi_k;t\in[0,T] \\ |\alpha|=1 \end{subarray}}
   \left[\partial^{\alpha} \Op H\left(\Delta\vphi_k,t\right)\right]\right]
  + \sup_{\begin{subarray}{c} \{\Delta\vphi_k\};t\in[0,T] \\ |\alpha|=2 \end{subarray}}
   \left[\partial^{\alpha}g\left(\{\Delta\vphi_k\},t\right)\right] \,,
\end{eqnarray}
\end{widetext}
i.e., $C$ is given by the supremum of
the first order derivatives of the Hamiltonian multiplied by the
norm of the costates $\chi_k(t)$ and the supremum of the first order
derivatives of the constraint, $g$.
For linear equations
of motion and $g_b$ zero or linear in $\vphi_k(t)$, we find $C=0$.
The case of a quadratic dependency of $g$ on the states $\vphi_k$, 
cf. Eq.~\eqref{eq:g_b}, can also easily be handled. 
The second term
in the right-hand side of Eq.~\eqref{eq:Cfinal} 
can then be estimated by the eigenvalue of the operator $\Op{D}(t)$
with largest magnitude. For example if $\Op{D}(t)$ is the
projector onto some subspace, $\Op{D}(t)=\Op{P}$,  then
\begin{equation}
  \label{eq:Cquad}
C \le -\frac{\lambda_b}{NT}\max_{\text{EV}}[\Op{P}]= -\frac{\lambda_b}{NT}\,.
\end{equation}

In order to derive a numerical approximation for the parameters $A$,
$B$ and $C$, we assume a finite time grid,
$\{t_{j}\}$, $j=1,\dots,n$, and define, based
on Eqs.~\eqref{eq:defA}, \eqref{eq:defB} and \eqref{eq:defC},
parameters $A^{(i+1)}$, $ B_j^{(i+1)}$ and $ C_j^{(i+1)}$,
\begin{widetext}
\begin{eqnarray}
  \label{eq:A_n}
  A^{(i+1)}(\{\Delta\vphi_k\})  & =&
  \frac{\sum_{k=1}^{N}\left[\Braket{\chi_k\left(T\right)|\Delta\vphi_k\left(T\right)}
    +\Braket{\Delta\vphi_k\left(T\right)|\chi_k\left(T\right)}\right]
    +J_T\left(\{\varphi_k^{(i)}(T)+\Delta\vphi_k(T)\}\right)
    -J_T\left(\{\varphi_k^{(i)}(T)\right)}
  {\sum_{k=1}^{N}\left[\Braket{\Delta\vphi_k\left(T\right)|\Delta\vphi_k\left(T\right)}\right]} \,,
\end{eqnarray}
and
\begin{eqnarray*}
  B_j^{(i+1)}(\{\Delta\vphi_k\}) &=& 
  \frac{\sum_{k=1}^{N}\left[\Braket{\Delta\vphi_k\left(t_j\right)|
        \Delta f_k\left(\Delta\vphi_k,t_j\right)}
      +\Braket{\Delta f_k\left(\Delta\vphi_k,t_j\right)|
        \Delta\vphi_k\left(t_j\right)}\right]}{\sum_{k=1}^{N}
    \left[\Braket{\Delta\vphi_k\left(t_j\right)|\Delta\vphi_k 
       \left(t_j\right)}\right]} \\ 
  C_j^{(i+1)}(\{\Delta\vphi_k\}) &=& 
  \frac{1}  {\sum_{k=1}^{N}\left[\Braket{\Delta\vphi_k\left(t_j\right)|\Delta\vphi_k
        \left(t_j\right)}\right]}  \bigg[
  \sum_{k=1}^{N}\big[\Braket{\dot{\chi}_k\left(t_j\right)|\Delta\vphi_k\left(t_j\right)}
    +\Braket{\Delta\vphi_k\left(t_j\right)|\dot{\chi}_k\left(t_j\right)} +
    \\ && \quad\quad\quad
    \Braket{\chi_k\left(t_j\right)|\Delta
      f_k\left(\Delta\vphi_k,t_j\right)}
    +\Braket{\Delta f_k\left(\Delta\vphi_k,t_j\right)|\chi_k\left(t_j\right)}\big]
    -\Delta g\left(\{\Delta\vphi_k\},t_j\right) \bigg]
\end{eqnarray*}
\end{widetext}
At iteration step $(i+1)$, a numerical estimate for the second order
parameter, $\sigma^{(i+1)}\left(t\right)$, is obtained by
replacing $A$, $B$, and $C$ in Eqs.~\eqref{eq:bar} by
$A^{(i+1)}$, Eq.~\eqref{eq:A_n}, together with
\begin{subequations}\label{eq:sigma_num}
\begin{eqnarray}
B^{(i+1)} &=& \sup_j B_j^{(i+1)}\,,   \label{eq:B_j} \\
C^{(i+1)} &=& \inf_j C_j^{(i+1)}\,,     \label{eq:C_j}
\end{eqnarray}
\end{subequations}
and inserting the resulting numerical estimates of $\bar{A}$,
$\bar{B}$, and $\bar{C}$ into Eqs.~\eqref{eq:sigma}.
Since the new states
$\vphi_k^{(i+1)}(t)=\vphi_k^{(i)}(t)+\Delta\vphi(t)$
are not known, $A^{(i+1)}$, $B^{(i+1)}_{j}$ and $C^{(i+1)}_{j}$ need to
be approximated, for example by the values of $A^{(i)}$,
$B^{(i)}_{j}$ and $C^{(i)}_{j}$ calculated in the previous 
iteration. In rare cases, this approximation might lead to loss
of monotonic convergence. The iteration then 
needs to be repeated with the values of $A^{(i+1)}$, $B^{(i+1)}_{j}$ and
$C^{(i+1)}_{j}$ that were obtained during the failed step.
Numerical estimation of the second order parameters
enforces monotonic convergence with respect
to a change in the states as gently as possible, making use of the
optimization history  to find a compromise between monotonicity and
speed of the convergence. 

\subsection{Monotonic convergence for arbitrary dependence of the
  Hamiltonian on the control}
\label{subsec:field}

In any iterative optimization, convergence with respect to the field
can only be expected towards a local extremum. Here, the local extremum condition on $J$
translates into $\frac{\partial^2 R}{\partial \epsilon^2} < 0$, or,
\begin{widetext}
  \begin{eqnarray}
    \frac{\partial^{2}g_a}{\partial\epsilon^{2}}
    \bigg|_{\vphi_{k}^{(i+1)},\epsilon^{(i+1)}} &>&
    \sum_{k=1}^{N}\Braket{\chi_{k}^{(i)}(t)\Bigg|
      \frac{\partial^{2}\Op H}{\partial\epsilon^{2}}
      \bigg|_{\epsilon^{(i+1)},\vphi^{(i+1)}}  \Bigg|\vphi^{(i+1)}_{k}(t)}
    +\sum_{k=1}^{N}\Braket{\vphi^{(i+1)}_{k}(t)\Bigg|
      \frac{\partial^{2}\Op H^{\dagger} }{\partial\epsilon^{2}}
      \bigg|_{\epsilon^{(i+1)},\vphi^{(i+1)}} 
      \Bigg|\chi_{k}^{(i)}(t)} \nonumber \\ &&+
    \frac{1}{2}\sigma(t)\sum_{k=1}^{N}\left[
      \Braket{\Delta\vphi_{k}(t) \Bigg|
        \frac{\partial^{2}\Op H}{\partial\epsilon^{2}}
        \bigg|_{\epsilon^{(i+1)},\vphi^{(i+1)}}
        \Bigg|\vphi_{k}^{(i+1)}(t)}
      +\Braket{\vphi_{k}^{(i+1)}(t)\Bigg|
        \frac{\partial^{2}\Op H^{\dagger}}{\partial\epsilon^{2}}
        \bigg|_{\epsilon^{(i+1)},\vphi^{(i+1)}} 
        \Bigg|\Delta\vphi_{k}(t)}\right]
    \,.
  \label{eq:locmaxReps}
\end{eqnarray}
\end{widetext}
For a linear dependence of the Hamiltonian on the control,
$\frac{\partial^{2}\Op H}{\partial\epsilon^{2}}=
\frac{\partial^{2}\Op H^{\dagger}}{\partial\epsilon^{2}}=
0$, and a maximum in $R$ requires simply 
$\frac{\partial^{2}g_a}{\partial\epsilon^{2}}\big|_{\epsilon^{(i+1)}} >
  0$.
This translates into the sign of the weight $\lambda_a$
for the typical quadratic dependence of $g_a$ on $\epsilon$,
cf. Eq.~(\ref{eq:g_a}). Inserting the corresponding derivative of
$g_a$, we obtain 
\begin{widetext}
\begin{eqnarray}
  \label{eq:eps1}
  \epsilon^{(i+1)}\left(t\right) &=& \tilde{\epsilon}\left(t\right)
  +\frac{S(t)}{\lambda_a }\mathfrak{Im}\Bigg\{ 
  \sum_{k=1}^{N}
  \Braket{\chi^{(i)}_{k}(t)\Bigg|
    \frac{\partial\Op H}{\partial\epsilon}
    \bigg|_{\epsilon^{(i+1)},\vphi^{(i+1)}} 
    \Bigg|\vphi_{k}^{(i+1)}(t)}
  + \frac{1}{2}\sigma(t)  \sum_{k=1}^{N}
  \Braket{\Delta\vphi_{k}(t)\Bigg|
  \frac{\partial\Op H}{\partial\epsilon}
  \bigg|_{\epsilon^{(i+1)},\vphi^{(i+1)}} 
  \Bigg|\vphi_{k}^{(i+1)}(t)}\Bigg\}\,.
\end{eqnarray}
\end{widetext}
For a non-linear dependency of the Hamiltonian on the
control, we define the change in the intermediate-time functional
due to changes in the control, 
\begin{eqnarray}
  \Delta_\epsilon(t) &=&
  R(\{\vphi_k^{(i+1)}(t)\},\epsilon^{(i+1)}(t),t)
  \nonumber \\ &&
  - R(\{\vphi_k^{(i+1)}(t)\},\epsilon^{(i)}(t),t) \,.
  \label{eq:Delta3}
\end{eqnarray}
The strict maximum condition for $R$ becomes
$\Delta_\epsilon(t) > 0$ $\forall t$.
We assume  $J_t$, cf. Eq.~(\ref{eq:J_t}), to be additive. 
Rewriting
$\Op H[\vphi_k^{(i+1)},\epsilon^{(i+1)},t]-\Op H[\vphi_k^{(i+1)},\epsilon^{(i)},t]$
in terms of the Taylor expansion of the Hamiltonian with respect to
the control, we find the zeroth order term to vanish.
The first order derivative can be rewritten using
Eq.~\eqref{eq:newfield}. The remaining terms correspond to 
the Taylor series starting at second order which can be estimated 
by its Lagrange remainder,
\begin{eqnarray}
  \label{eq:Mtilde}
\tilde{M}^\epsilon_2(t) = \sup_{\varphi_k;\epsilon}
\left|\frac{\partial^2}{\partial\epsilon^2}
  \Op H(\varphi_k,\epsilon,t)\right|\,. 
\end{eqnarray}
Employing furthermore
$\sum_{k=1}^{N}\sqrt{\Braket{\Delta \vphi_k(t)|\Delta
      \vphi_k(t)}} \leq 2\sqrt{N}$,
we obtain
\begin{widetext}
\begin{eqnarray*}
   \Delta_\epsilon(t) & > &
   g_a(\epsilon^{(i)},t)
   -g_a(\epsilon^{(i+1)},t)
   + \left(\epsilon^{(i+1)}(t)-\epsilon^{(i)}(t)\right)
   \frac{\partial g_a}{\partial\epsilon}\bigg|_{\epsilon^{(i+1)}}
  \\ &&
   +\Bigg\{
   \sqrt{N}   \tilde{M}^\epsilon_2(t) 
   \sum_{k=1}^{N}\sqrt{\Braket{\chi_k^{(i)}(t)|\chi_k^{(i)}(t)}}
   +|\sigma(t)|N \tilde{M}^\epsilon_2(t)
   \Bigg\}
   \left(\epsilon^{(i+1)}(t)-\epsilon^{(i)}(t)\right)^{2}
   \,,
\end{eqnarray*}
\end{widetext}
The Lagrange remainder can be evaluated by taking the second derivative of the
Hamiltonian with respect to the field and estimating the norm of the
resulting  operator by its spectral radius or its eigenvalue with
largest square modulus. 
Since it
is difficult to proceed without a specific dependence of $g_a$ on
$\epsilon$, we assume a quadratic dependence, cf. Eq.~(\ref{eq:g_a}), and find
\begin{widetext}
\begin{eqnarray}
  \label{eq:dRgen}
  \Delta_\epsilon(t) & > & \left[
    \frac{\lambda_a}{S(t)}-\left\{\frac{1}{2} \sqrt{N}
      \sum_{k=1}^{N}\left[\sqrt{\Braket{\chi_k^{(i)}(t)|\chi_k^{(i)}(t)}}\right]
      \tilde{M}^\epsilon_2(t)
      +N |\sigma(t)| \tilde{M}^\epsilon_2(t)\right\}
    \right] \left(\epsilon^{(i+1)}(t)-\epsilon^{(i)}(t)\right)^{2} \,.
  \end{eqnarray}
Monotonic convergence is ensured by adjusting the shape function
$S(t)$ and the parameter $\lambda_a$ such that 
\begin{eqnarray}
  \label{eq:lambda}
  \frac{\lambda_a}{S(t)}  & > &
  \bigg\{\frac{1}{2} \sqrt{N} \sum_{k=1}^{N}\left[
    \sqrt{\Braket{\chi_k^{(i)}(t)|\chi_k^{(i)}(t)}}\right]
      \tilde{M}^\epsilon_2(t) 
      +N |\sigma(t)| \tilde{M}^\epsilon_2(t)\bigg\}\,.
\end{eqnarray}
\end{widetext}
Unlike the algorithm of Ref.~\cite{OhtsukiPRA08}, the
numerical effort in our approach to ensure monotonic convergence
is independent of the order of the non-linearity of the Hamiltonian's
dependence on the control.

To summarize section~\ref{sec:OCT}, 
a second order construction of the optimization
algorithm yields an additional contribution to the equation for the
new field. Compared to the linear variant of Krotov's
method~\cite{JosePRA03}, it simply requires additional 
storage of the states from
the previous iteration to determine $\Delta\vphi_k(t)$
and calculation of the function $\sigma(t)$.
The parameters $A$, $B$ and $C$ determining $\sigma(t)$
turn out to be zero, i.e., the second order contribution
vanishes, for the 'standard' quantum control problem with bilinear
final-time cost, intermediate-time costs that are independent of the
states and linear equations of motion. A second order
contribution can nevertheless be invoked to study its influence on
convergence.  
This is investigated below in Section~\ref{subsec:res_1}. 
A second order contribution is required to
guarantee monotonic convergence for final-time costs that are a
polynomial of higher than second order in the states~\cite{Matthias},
demonstrated below in Section~\ref{sec:J_LI},
and for general intermediate-time costs that depend on the states,
studied below in Section~\ref{subsec:res_2}. A second order
construction is also required for non-unitary time evolution and for
non-linear equations of motion~\cite{SklarzPRA02,MundtNJP09}.


\section{Application I: Higher order polynomial costs}
\label{sec:J_LI}

A functional that is an eighth order polynomial in the states
arises in quantum information when optimizing 
for a local equivalence class of two-qubit gates, $[\Op O]$, rather
than a specific two-qubit gate, $\Op O$~\cite{Matthias}. The
functional is based on the local invariants of two-qubit
gates~\cite{ZhangPRA03} which 
uniquely characterise a local equivalence class.
The explicit, somewhat lengthy expression of $J^{LI}_T$
is given in Ref.~\cite{Matthias}. 

We employ $J^{LI}_T$  to optimize for the local equivalence class of the
B-gate~\cite{ZhangPRL04}, given in the logical basis by 
\begin{eqnarray*} 
\Op O_{B} &=& 
e^{\frac{i}{2}\left[\frac{\pi}{2}\Op\sigma_x\otimes\Op\sigma_x +\frac{\pi}{4}\Op\sigma_y\otimes\Op\sigma_y\right]} \\
&=& \begin{pmatrix}
  \cos(\pi/8) & 0 & 0 & i \sin(\pi/8) \\
  0 & \sin(\pi/8) & i\cos(\pi/8) & 0 \\
  0 & i\cos(\pi/8) & \sin(\pi/8) & 0 \\
  i\sin(\pi/8) & 0 & 0 & \cos(\pi/8) 
\end{pmatrix}\,,
\end{eqnarray*}
for an effective spin-spin system,
\begin{equation}
  \label{eq:Heff}
\Op H_{eff} = \frac{\hbar\Omega(t)^2}{8}\sum_{i,j=0}^3 \Op\sigma_i 
a_{ij}(x_0) \Op\sigma_j\,.
\end{equation}
The effective Hamiltonian is derived, within second-order perturbation
theory, for trapped polar molecules with $^2\Sigma_{1/2}$ electronic
ground  states, subject to near-resonant microwave driving that induces
strong dipole-dipole coupling \cite{MicheliNatPhys06}. $\Omega(t)$
denotes the Rabi frequency that will be optimized, $\Op\sigma_i$ are
the $2\times 2$ Pauli spin matrices, $i=1,2,3 \equiv x,y,z$ with $\sigma_0=\openone_2$. 
The tensor $a_{ij}$ depends on the
distance $x_0$ between the molecules and on the polarization and
detuning of the microwave field. We consider here CaCl molecules 
in an optical lattice with a lattice spacing of $300\,$nm, microwave
radiation of about $9.13\,$GHz, polarizations
$\alpha_{\pm}=1/\sqrt{2}$, $\alpha_0=0$, and a detuning from the
rotational transition of $1.2\,$kHz.

\begin{figure}[tb]
  \centering
  \includegraphics[width=0.9\linewidth]{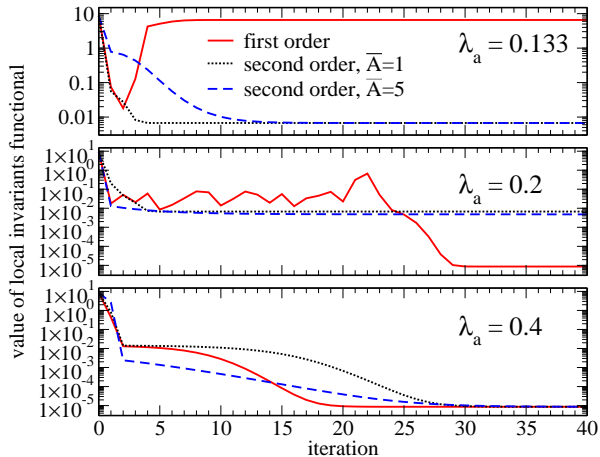}
  \caption{(color online)
    Convergence of the local invariants
    functional that optimizes for a local equivalence class rather
    than a specific operator and is an eighth order polynomial in the
    states. The optimum corresponds to $J^{LI}_T=0$.} 
  \label{fig:CaCl}
\end{figure}
The higher than quadratic dependence of $J^{LI}_T$ on the states leads
to a non-zero analytical estimate 
of $A$, cf. Eq.~\eqref{eq:A_est}. Specifically, one has to calculate
all second derivatives of the eighth order polynomial and
evaluate the supremum by considering an arbitrary change in the states
up to $\|\Delta \varphi\| = \|\sum_k \Delta \varphi_k\| \le 2\sqrt{N}$,
cf. Appendix~\ref{sec:adaptproof} (Fig.~\ref{fig:norm}). For our
example, $N=4$ and from the second derivatives, we obtain $\bar{A}=45$
although in practice a value of $\bar{A}=5$ was sufficient to preserve
monotonic convergence.
Figure~\ref{fig:CaCl} demonstrates that this choice of $\bar{A}$ (blue
dashed line) indeed 
ensures monotonic convergence independent of other parameters of the
algorithm such as the weight $\lambda_a$, cf. Eq.~\eqref{eq:g_a}. 
Note that in this example convergence of the final-time functional
$J_T$ and the complete functional $J$ are equivalent due to our
specific choice of $g_a$~\cite{JosePRA03} and $g_b=0$. The latter
yields $\bar{C}=0$. Furthermore $\bar{B}=0$ since 
our equation of motion is linear in the states.  
The first order algorithm (solid red line) fails completely for
small $\lambda_a$ and violates monotonic convergence for many
iteration steps albeit finding an optimum eventually for intermediate
$\lambda_a$. The weight $\lambda_a$ determines the step size for
changes in the field, cf. Eq.~(\ref{eq:neweps}):
small $\lambda_a$ leads to large values of the field and thus a
possibly more 'erratic' optimization, while for large $\lambda_a$,
optimization proceeds more cautiously, explaining that
even the first order construction is found to converge
monotonically. This is, however, due to the 
simplicity of our control problem which is essentially one-dimensional
since success is determined by the integrated field amplitude. For
more complex Hamiltonians, optimization employing a first order
construction was found to always fail~\cite{Matthias}, underlying the
significance of the second order construction. 
The analytical estimate  of the second order parameter 
takes all worst-case scenarios into account. Therefore values smaller
than the analytical estimate 
might already be sufficient for monotonicity which is confirmed 
as shown in Fig.~\ref{fig:CaCl} by the blue dashed line representing $\bar{A}=5$ and
the black dotted line representing $\bar{A}=1$ even though the analytical estimate is
given by $\bar{A}=45$. 
Such a less conservative estimate yields, moreover, 
signficantly faster convergence for small $\lambda_a$, a fact that 
was also confirmed for more complex Hamiltonians~\cite{Matthias}. 

\section{Application II: Bilinear costs}
\label{sec:appl}

For bilinear costs, the supremum estimation of $C$ yields zero. We
consider the dynamics of our examples to be described by the standard
Schr\"odinger equation such that $B=0$ and $\bar{B}$ can also be
chosen zero. A second-order
construction of the optimization algorithm becomes necessary if
bilinear intermediate-time costs are employed, $g_b \neq 0$, 
which lead to non-zero $C$, cf. Eq.~\eqref{eq:Cfinal}. If $g_b = 0$, a
second-order construction is not required to ensure monotonicity but 
can be utilized to speed up convergence. 

\subsection{Model}
\label{subsec:model}

A simplified model for the vibrations in an Rb$_2$ molecule 
that linearly interacts with a laser field  takes
three electronic states, $X^1\Sigma_g^+(5s+5s)$, $^1\Sigma_u^+(5s+5p)$
and $^1\Pi_g(5s+4d)$, 
into account \cite{JoseMyPRA08,NdongMyJCP09}, 
\begin{eqnarray}
  \Op{H}= \sum_{i=1}^{3}\Op{H}_i\otimes |e_i\rangle\langle e_i| +
  \Op{\mu}\,\epsilon(t) \cdot \,\big(|e_1\rangle\langle e_2| &&
  \nonumber    \\
  + |e_2\rangle\langle e_1|+
  |e_2\rangle\langle e_3|+|e_3\rangle\langle
  e_2| \big)\,. &&
  \label{eq:H}
\end{eqnarray}
Here, $\Op{H}_i$ denotes the vibrational Hamiltonians,
$\Op{H}_i=\Op{T}+V_i(\Op{R})$, with kinetic and potential operators
$\Op{T}$ and $V_i(\Op{R})$, respectively, 
$\Op{\mu}$ is the transition dipole operator, assumed to be
independent of $\Op{R}$, and $\epsilon(t)$ the
electric field. The potentials are found in
Ref.~\cite{ParkJMS01}. The vibrational Hamiltonians are represented on
a Fourier grid \cite{RonnieReview88} with $N_R$ grid points yielding a
total Hilbert space dimension
of $M=3 N_R$. The equations of motion are solved by the Chebychev
propagator for homogeneous and inhomogeneous Schr\"odinger equations,
respectively \cite{RonnieReview88,NdongMyJCP09}.
An initial field of the form
\begin{equation}
  \epsilon^{(0)}(t) = \epsilon_0s(t)\cos(\Omega t)
\end{equation}
is employed with $\epsilon_0$ the maximum amplitude 
and $\Omega$ the central frequency of the field. The shape function $s(t)$
is taken to be $s(t) = \sin^2(\pi t/T)$, where $T$ corresponds to
the optimization time. The weight of the final-time objective,
$\lambda_0$,  is taken to be one such that the optimum corresponds to
$-J_T=1$. 

\subsection{State-independent  intermediate-time cost ($g_b=0$)}
\label{subsec:res_1}

We investigate optimization of a state-to-state transfer
($N=1$), taking for simplicity only the electronic states
$X^1\Sigma_g^+(5s+5s)$ and $^1\Sigma_u^+(5s+5p)$ into account.
The initial and target states are taken to be the vibrational
eigenstates $v=10$ and $v=0$ of the $X^1\Sigma_g^+(5s+5s)$ electronic 
ground state. With a vibrational period of the initial state of
614$\,$fs, the optimization time is set to $T = 1\,$ps. The central
frequency and maximum field amplitude are
taken to be $\Omega = \omega_{v=10\rightarrow v'=0}$ and $\epsilon_0 =
1\cdot 10^{-2}\,$a.u. 

\begin{figure}[tb]
\includegraphics[width = 0.9\linewidth]{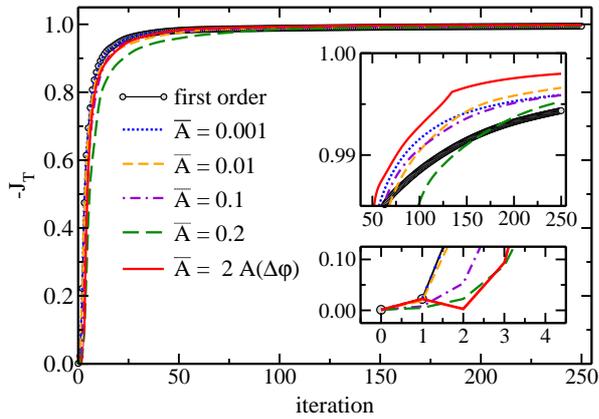}
\caption{(color online)
  Convergence of the first order and second order constructions of the
  optimization algorithm  as
  measured by the final-time objective, $J_T$,
  for state-to-state transfer from vibrational level $v=10$ to $v=0$.
}
\label{fig:JTstate2state}
\end{figure}

Convergence of the final-time objective $J_T$ is shown in
Fig.~\ref{fig:JTstate2state}, comparing first order (black circles)
and second order constructions of the algorithm.
The second order construction is determined by the choice of 
$\bar{A}$ which can be taken to be equal to some non-negative
number, $\varepsilon_A$, cf. Eq.~(\ref{eq:bar}) (dotted and dashed lines in
Fig.~\ref{fig:JTstate2state}) or $\bar{A}=0$ (first order
optimization, black circles in Fig.~\ref{fig:JTstate2state}). The
numerical estimate of $A$, cf. Eq.~(\ref{eq:A_n}), is represented by
the solid red line in Fig.~\ref{fig:JTstate2state}.
The latter choice might speed up convergence, but is more risky: Since
$\bar{A}=2A^{(i+1)}(\Delta\vphi)$ can become negative, the condition for
monotonic convergence may be violated. This is clearly seen in
Fig.~\ref{fig:JTstate2state}. In the lower inset, monotonic
convergence is lost for one step after the first iteration step. We
find in this case, that the state change is almost maximal,
$\|\Delta\vphi\|=1.95\le 2$, i.e., the worst possible case that the
optimization algorithm must deal with is reached. While the
first order construction converges faster initially, the upper inset
shows that all second order constructions supersede the first order
one as the optimum is approached. This is readily understood by
inspection of Eq.~(\ref{eq:neweps}): The first order contribution to
the change in the field is closely related to the gradient of the
functional.~\footnote{
  Specifically, the gradient in an implementation of the
  GRAPE algorithm employing a sequential update of the field 
  coincides with the first order term of
Eq.~\eqref{eq:neweps}~\cite{KhanejaJMR05}.}  
Since the gradient 
vanishes close to the optimum, convergence of the first order
construction slows down as the optimum is approached. 
Variation of the non-negative
number, $\varepsilon_A$, shows that an optimal choice of
$\varepsilon_A$ exists. However, this optimal choice cannot be
determined \textit{a priori}. In terms of convergence speed close to
the optimum, it is therefore recommendable to employ the numerical
estimate of $\bar{A}$ (red solid line in
Fig.~\ref{fig:JTstate2state}.). 
Very similar behavior is found for optimization of a unitary
transformation, a Hadamard gate ($N=2$) on the lowest
two vibrational eigenstates of the electronic ground state (data not
shown).

\subsection{State-dependent  intermediate-time cost ($g_b\neq 0$)}
\label{subsec:res_2}

State-to-state transfer from $v=0$ to
$v=1$ and the Hadamard gate  on the lowest
two vibrational eigenstates of the electronic ground state,
\[
\Op{O}
= \frac{1}{\sqrt{2}}
\begin{pmatrix}
1 & 1 \\
1 & -1
\end{pmatrix} \,,
\]
are optimized, taking into
account an additional state-dependent cost, 
$g_b\left(\vphi,t\right)$. 
If both a state-dependent
cost and a final-time target are present, the algorithm seeks to
optimize a compromise between the two goals. The parameters
$\lambda_0$ and $\lambda_b$ determine the relative weight of each
target. Monotonic convergence always refers to the value that the
total functional, $J$ of Eq.~(\ref{eq:J}), takes; and each separate
contribution to $J$ does not need to converge monotonically. Below we
will discuss convergence of both $J$ and $J_T$. In order to render the
optimal value of $J$ independent of the choice of the weights
$\lambda_0$, $\lambda_b$, we define a normalized functional, 
\begin{subequations}
  \begin{eqnarray} 
    J_{norm} &=&   \frac{J}{\lambda_b-\lambda_0},  \quad \lambda_b \le 0\,,
    \label{eq:J_normallowed}\\
    J_{norm} &=&   1 - \frac{J-\lambda_0}{\lambda_b-\lambda_0},  \quad \lambda_b \ge 0\,,
    \label{eq:J_normforbi}
  \end{eqnarray}
\end{subequations}
that converges toward one.

The cost $g_b$ is employed to avoid any population transfer to a
forbidden subspace, taken to be the  $^1\Pi_g(5s+4d)$ state,
at all times  $t\in [0,T]$ \cite{JoseMyPRA08}.
This can be expressed by taking the operator $\Op{D}$ in
Eq.(\ref{eq:g_b}) to be one of the two choices,
\begin{subequations}
\begin{eqnarray}
\Op{D} &=& \Op{P}_{\mathrm{allow}} = |e_1\rangle\langle e_1| +
           |e_2\rangle\langle e_2|,  \quad \lambda_b \le 0
\label{eq:Dallowed}\\
\Op{D} &=&   \Op{P}_{\mathrm{forbid}}  =
           |e_3\rangle\langle e_3|,  \quad \lambda_b \ge 0\,,
\label{eq:Dforbid}
\end{eqnarray}\label{eq:Ds}
\end{subequations}
where $ \Op{P}_{\mathrm{allow}} $ and $\Op{P}_{\mathrm{forbid}}$ denote the
projectors onto the allowed and forbidden subspaces, respectively.
The allowed subspace is formed by the $X^1\Sigma_g^+(5s+5s)$ and
$^1\Sigma_u^+(5s+5p)$ states. 
The different signs of the weight $\lambda_b$ indicate maximization
of $\Op{P}_{\mathrm{allow}}$ and minimization of
$\Op{P}_{\mathrm{forbid}}$ which is physically equivalent. 
Mathematically, the equivalence does not hold since $g_b$
is a constraint and therefore must have its sign opposite to that of
$J_T$.  This is possible only for the choice of
Eq.~\eqref{eq:Dallowed}. Note that the 'wrong' sign of
Eq.~\eqref{eq:Dforbid} exemplifies the more general
class of indefinite operators $\Op{D}(t)$ for which it is not possible
to construct a monotonically convergent algorithm using the previously
available tools by a simple change of sign.

With the choice of $g_b$ according to Eqs.~\eqref{eq:Ds}, the
analytical estimate of  
the parameter $C$ of the second order contribution is
given by Eq.~(\ref{eq:Cquad}). 
Writing explicitly the change in the intermediate-time contribution to
the functional due to the change in the states for a first order
algorithm, 
\begin{eqnarray*}
\Delta_{\vphi(t)} &=& R(\{\vphi_k^{(i+1)}\},\epsilon^{(i)},t) -
 R(\{\vphi_k^{(i)}\},\epsilon^{(i)},t)  \\
&=& -\lambda_b\;\frac{1}{NT}\sum_{k=1}^N
\langle \Delta\vphi_k(t)| \Op{D}|
\Delta\vphi_k(t) \rangle \,.
\end{eqnarray*}
we find  the necessary condition for monotonic convergence,
$\Delta_{\vphi(t)} \ge 0$, to be always fulfilled
for $\Op{D}=\Op{P}_{\mathrm{allow}}$. A second order construction is
therefore not required~\cite{JoseMyPRA08}, corresponding to
$C=0$ in 
accordance with Eq.~(\ref{eq:Cquad}). In this case, 
$\bar{C}$ can be set to $-\varepsilon_C$,
cf. Eq.~(\ref{eq:bar}), where  $\varepsilon_C$ is a non-negative
number to check for improved convergence with a second order
algorithm. However, if the projector onto the forbidden
subspace is employed in $g_b$, $\Delta_{\vphi(t)} \ge 0$ is not necessarily
fulfilled  and a second order construction is
required to ensure monotonic convergence.
Eq.~(\ref{eq:Cquad}) now yields a large negative number for $C$
since $\lambda_b$ is negative, and  $\bar{C}$ is determined by $C$. 
Our approach goes beyond the
results of Ref.~\cite{JoseMyPRA08} where a convergent (first order)
algorithm was obtained only for negative semi-definite operators
$\lambda_b\Op{D}(t)$ in the additional constraint $g_b$. 
The second order construction of Eq.~\eqref{eq:eps1} allows for a
larger class of operators $\Op{D}(t)$ in the state-dependent
constraint $g_b$.

The final time is set to $T=2\,$ps, the central frequency  
of the guess field is chosen to be $\Omega =
\omega_{v=0\rightarrow v'=10}$
and $\epsilon_0 = 2\cdot 10^{-4}\,$a.u.
$\bar{A}$ is taken to be zero since our emphasis is on the choice of
the parameter $\bar{C}$. 

\begin{figure}[tb]
  \includegraphics[width = 0.9\linewidth]{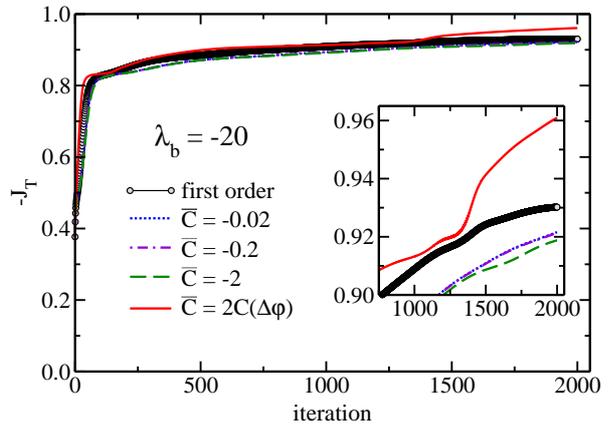}
  \caption{(color online)
    Convergence of the first order and second order constructions for
    a Hadamard gate with a state-dependent cost
    ($\Op{D}=\Op{P}_\mathrm{allow}$, 
    i.e. the second order construction is not required).
  }
  \label{fig:JT_U_allowlb20}
\end{figure}

In analogy to the previous subsection, we first study the case
$\Op{D}=\Op{P}_\mathsf{allow}$ where the second order construction is
not required but may improve convergence.
Figure~\ref{fig:JT_U_allowlb20} compares the convergence of
first and second order constructions.
Taking $\bar C$ to be equal to $-\varepsilon_C$,
cf. Eq.~(\ref{eq:bar}) (dotted and dashed lines in 
Fig.~\ref{fig:JT_U_allowlb20}), does not affect
monotonicity. However, it also does not yield faster convergence
than the first order algorithm (black circles). 
The numerical estimate, $\bar C = 2 C^{(i+1)}(\Delta\vphi)$ (solid red
line in Fig.~\ref{fig:JT_U_allowlb20}), cf.
Eq.~(\ref{eq:C_j}), neglecting $\varepsilon_C$ in Eq.~(\ref{eq:bar})
is somewhat risky since $C^{(i+1)}(\Delta\vphi)$ can
become positive such that this choice of $\bar 
C$ does not guarantee monotonic convergence. 
Indeed, small violations of monotonicity, for
example between steps 1180 or 1300, are observed in
Fig.~\ref{fig:JT_U_allowlb20}.
However, this is more than compensated for by the improved speed of
convergence as compared to the first order and the conservative
choices $\bar C=-\varepsilon_C$. 
Very similar behavior is found for optimization of a state-to-state
transition (data not shown). 

\begin{figure}[tb]
  \includegraphics[width = 0.9\linewidth]{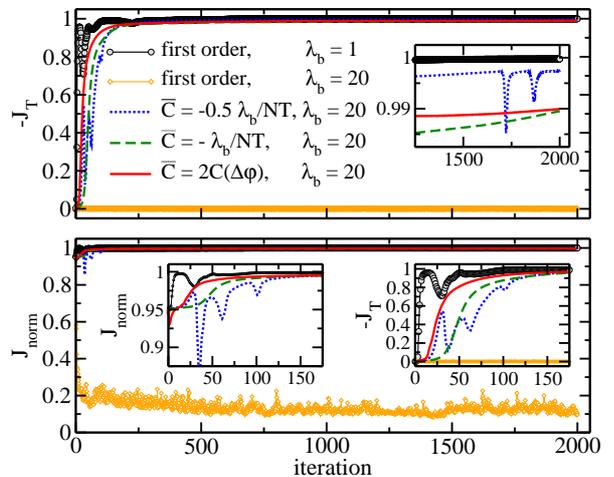}
  \caption{(color online)
    Convergence of the first order and second order constructions for
    state-to-state transfer with state-dependent cost.
    The operator $\Op{D}$ is taken to be the projector onto a forbidden
    subspace, i.e. the second order construction \textit{is} required.
  }
  \label{fig:JT_forbid20}
\end{figure}

For $\Op{D}=\Op{P}_{\mathrm{forbid}}$, monotonic convergence needs to
be ensured by a second order construction with $C$ given by 
Eq.~(\ref{eq:Cquad}). This choice corresponds to the green dashed line
in Fig.~\ref{fig:JT_forbid20} which studies convergence of the
final-time objective, $J_T$, and the complete functional,
$J_\mathrm{norm}$,  for a state-to-state optimization. 
Taking $C$ somewhat smaller than the
estimate of Eq.~(\ref{eq:Cquad}) may yield non-monotonic behavior,
cf. the blue dotted line in Fig.~\ref{fig:JT_forbid20}.
If one neglects the second order contribution and the weight
$\lambda_b$ is large, the algorithm completely fails (orange diamonds
in Fig.~\ref{fig:JT_forbid20}). For  a small weight $\lambda_b$, 
the algorithm converges to an optimum but non-monotonic behavior 
is observed at intermediate iterations (black circles). 
Note that for small $\lambda_b$, the constraint
$g_b$ is almost not enforced due to insufficient weight.
The best compromise between monotonic convergence and high fidelity is
obtained for $\bar C = 2 C^{(i+1)}(\Delta\vphi)$ (red solid line in
Fig.~\ref{fig:JT_forbid20}).
For the parameters for which the complete functional,
$J_\mathrm{norm}$, converges monotonically
(green dashed and solid red lines in Fig.~\ref{fig:JT_forbid20}),
monotonic behavior is observed also for the final-time objective,
$J_T$. In general, this need
not be the case. We attribute it in the current example to our choice
of the guess field which is relatively weak such that the 
forbidden subspace is not strongly populated. The algorithm
therefore starts out in the 'right' direction for optimizing both
targets, $J_T$ and $g_b$, and it does not need to optimize one target
at the expense of the other.  

\begin{figure}[tb]
  \includegraphics[width = 0.9\linewidth]{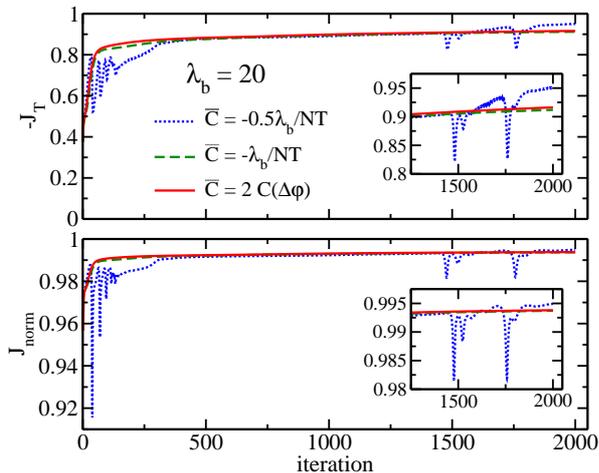}
  \caption{(color online)
    Convergence of second order constructions for
    a Hadamard gate with a state-dependent cost.
    The operator $\Op{D}$ is taken to be the projector onto a forbidden
    subspace, i.e. the second order construction \textit{is} required.
  }
  \label{fig:JT_U_forbidlb20}
\end{figure}

Fig.~\ref{fig:JT_U_forbidlb20} presents convergence of the
final-time objective, $J_T$, and the complete functional,
$J_\mathrm{norm}$,  for optimization of the Hadamard gate.
Similarly to Fig.~\ref{fig:JT_forbid20},  convergence is almost
identical for the analytical estimate of $C$ based on
Eq.~(\ref{eq:Cquad}) (green  dashed line in
Fig.~\ref{fig:JT_forbid20}) and the numerical estimate according to 
Eq.~(\ref{eq:C_j}) (red solid line). A small violation of
the analytical estimate (blue dotted line in
Fig.~\ref{fig:JT_U_forbidlb20}) leads to non-monotonic behavior but
may yield larger fidelities after many iterations.

To summarize our numerical investigations,  a second order
contribution can be employed to enforce monotonic convergence for
functionals that are higher-order polynomials in the states or
correspond to expectation values of non-semidefinite operators. 
The numerical estimate of the second order paramters might 
slightly violate monotonicity but yields the highest fidelities,
especially as the optimum is approached. If a
second order contribution is not required by the functional, 
it may nevertheless be used to improve convergence. Also in this case
the numerical estimate of the second order 
parameters $\bar A$ and $\bar C$ turns out to be most efficient.

\section{Summary and Conclusions}
\label{sec:concl}

Applying Krotov's method~\cite{Konnov99} to quantum control, we have shown
that monotonically convergent optimization algorithms are obtained for
\textit{any} quantum control problem provided that a second order
construction is employed. The equation for the optimized 
field then contains an additional term. Compared to a first order
algorithm~\cite{JosePRA03}, only storage of the quantum states of the
previous iteration and calculation of the second order weight are
additionally required. We have shown that the parameters for the
second order contribution can be estimated analytically 
based  on the final-time target and intermediate-time
'costs', the equations of motion and the dependence of the
Hamiltonian on the control field or calculated numerically from the
optimization history. This is due to the 
normalization of quantum state vectors and finiteness of physical
control fields, implying that optimization is performed over compact
sets of candidate states and controls, which has allowed us to 
significantly relax the conditions for Krotov's constructive
proof~\cite{Konnov99}. 

We have illustrated the power of our approach by applying it to two
control problems for which no monotonically convergent algorithm 
existed  -- to target functionals that are higher-order
polynomials in the states and to state-dependent constraints expressed 
as expectation value of a non-semidefinite operator. Target functionals
that are higher order polynomials in the states arise for optimization
towards an equivalence class of operators rather than a specific
operator. This is particularly relevant in quantum
information where one is interested in the optimal evolution of a
primary system alone, irrespective of its
environment~\cite{GraceNJP10}, or in the entangling content of
two-qubit gates~\cite{Matthias}.  

Our numerical examples illustrate that 
an analytical estimate of the algorithm parameters ensures
monotonic convergence by taking all
worst-case scenarios for optimization into account. However, if 
worst-case scenarios are not encountered, the analytical
estimate imposes limits which are too severe, slowing down
convergence. Estimating the algorithm parameters numerically based on 
the optimization history turns out to be a more efficient choice.
The numerical estimate of the second order parameters can also be
employed to speed up convergence, in particular close to the optimum
where the first order contribution vanishes,  for optimization
problems where a second order construction is not strictly required. 

Note that the overall performance of the algorithm still depends on the
the weights of the each term in
the optimization functional and the optimization time
$T$. The latter has little influence on the convergence once it is
larger than the quantum speed
limit~\cite{CanevaPRL09}. The role of the weights is more subtle, in
particular for multi-optimization problems~\cite{RabitzPRA08}. A
special role is taken by the weight, $\lambda_a$, of the term
minimizing the field intensity: It determines the magnitude of
the first order contribution to the new field analogously to the step
size in gradient-type algorithms~\cite{KhanejaJMR05}. Since its
modulus, $|\lambda_a|$, is a free parameter of the algorithm, its
choice may be used to further improve the convergence
speed. An efficient optimization method is obtained by choosing 
$|\lambda_a|$ based on information from the second order derivative of the
functional with respect to the field, estimated with the
Broyden-Fletcher-Goldfarb-Shanno (BFGS) algorithm~\cite{EitanPRA11}. 
Note that in terms of convergence speed
this goes beyond our approach which only makes use of
information from the second order derivatives with respect to the
states. 

The work presented here opens up a whole range of new applications for
quantum optimal control. It provides a general set of tools to study
optimization of final-time functionals that are
higher order polynomials in the states~\cite{Matthias}, or 
optimization of time-dependent expectation values that were suggested
for the control of high-harmonic emission~\cite{SerbanPRA05}, or 
optimization for non-linear equations of
motion such as the Gross-Pitaevski equation, time-dependent
Hartree-Fock equations or time-dependent density functional
theory~\cite{GrossPRL07,GrondPRA09,MundtNJP09}. 
This set of tools allows for designing novel optimization functionals 
that capture the relevant physics without
restriction to bilinear functionals. 


\begin{acknowledgments}
  We would like to thank Reuven Eitan and David Tannor for their
  helpful comments on the manuscript.
  Financial support  from the Deutsche Forschungsgemeinschaft
  is gratefully acknowledged.
\end{acknowledgments}

\appendix
\section{Proof of optimality of the second order ansatz}
\label{sec:proof}


Konnov and Krotov base their proof upon the following 
condition set $\mathfrak{A}$~\cite{Konnov99}:
\begin{enumerate}
\item The right-hand side of the equation of motion,
  $f\left(\vec{\vphi},\epsilon,t\right)$, is bounded, i.e.
  $\exists \; K,L<\infty: \forall \left(\vec{\vphi},\epsilon,t\right)
  \in\mathbb{R}^{2NM}\times\mathbb{E}\times\left[0,T\right],\|x\|\geq
  M\Rightarrow f\left(\vec{\vphi},\epsilon,t\right)\leq L\cdot\|x\|$.
\item
  The Jacobian of the equation of motion is bounded, i.e.
  $\exists \; A<\infty: \forall \left(\vec{\vphi},\epsilon,t\right)
  \mathbb{R}^{2NM}\times\mathbb{E}\times\in[0,T]:\|\boldsymbol{\mathsf{J}}\|\leq
  A$.
\item The functionals $J_T(\vec{\vphi})$ and
  $g\left(\vec{\vphi},\epsilon,t\right)$ are twice differentiable
  and bounded, i.e.
  $\exists\; K,L<\infty: \forall\vec{\vphi}\in\mathbb{R}^{2NM},
  \|\vec{\vphi}\|\geq L\Rightarrow J_T\left(\vec{\vphi}\right)\leq
  K\|\vec{\vphi}\|^{2}
  \text{ and }
  \left|g\left(\vec{\vphi},\epsilon,t\right)\right| \leq
  K\|\vec{\vphi}\|^{2} \,\forall \left(t,\epsilon\right)\in
  \left[0,T\right]\times\mathbb{E}$.
\end{enumerate}
Here, the real state vector $\vec{\vphi}(t)$ is a piecewise
differentiable function for all $t\in\left[0,T\right]$
and  the control $\epsilon$ is an element of
the Banach space of continous real valued functions on the interval
$\left[0,T\right]$ with supremum norm
$\|f\left(x\right)\|_{\infty}=\sup_{x\in\left[0,T\right]}\left|f\left(x\right)\right|$,
i.e. $\epsilon\in\left(C\left[0,T\right],\|\cdot\|_{\infty}\right)$
with $\|\epsilon\|_{\infty}\leq E<\infty$. 
In quantum control, the components of $\vec{\vphi}(t)$ 
are the real and imaginary parts of the projections of $N$ Hilbert
space vectors on a suitable orthonormal basis of the $M$-dimensional
Hilbert space. The norm of the Jacobian is
then the matrix norm of the $2NM\times 2NM$ matrix, either column or row norm.

Based on $\mathfrak{A}$, Konnov and Krotov proved the following
theorem~\cite{Konnov99}: 
\begin{thm}\label{thm:1}
  If the conditions $\mathfrak{A}$ hold, then for each process 
  $w^{(i)}\left(\vec{\vphi}^{(i)}(t),\epsilon^{(i)}(t)\right)$,
  there exists a solution for $\Phi$ to the extremization problem,
\begin{eqnarray}
  R\left(\vec{\vphi}^{(i)}(t),\epsilon^{(i)}(t),t;\Phi\right)
  & = & \min_{\vec{\vphi}}
  R\left(\vec{\vphi}(t),\epsilon^{(i)}(t),t\right)
  \nonumber \\
  && \quad\; \forall\; t\in\left[0,T\right]\,,\label{eq:minR} \\
  G\left(\vec{\vphi}^{(i)}(T);\Phi\right) & = &
  \max_{\vec{\vphi}} G\left(\vec{\vphi}(T)\right)\,,
\label{eq:maxG}
\end{eqnarray}
of  the form
  \[
  \Phi(\vec{\vphi},\vec{\chi},t) =
  \vec{\chi}\left(t\right)\cdot\vec{\vphi}\left(t\right)+
  \frac{1}{2}\Delta\vec{\vphi}(t)\cdot\Op{\Sigma}(t)\cdot\Delta\vec{\vphi}(t)\,,
  \]
  and the matrix function $\Op{\Sigma}(t)$
  can be represented as
  \[
  \Op{\Sigma}(t) =
  \left(\alpha\left(e^{\gamma\left(T-t\right)}-1\right)+\beta\right)\cdot
  \openone\equiv\sigma\left(t\right)\cdot \openone \,,
  \]
where $\alpha,\beta<0$ and $\gamma>0$ are constants.
\end{thm}
With this theorem,
the crucial part of constructing the algorithm, i.e. determination of
$\sigma(t)$ and thus $\Phi$, is  reduced to the determination of the
constants $\alpha$, $\beta$
and $\gamma$. Since the proof of Theorem~\ref{thm:1} shows how to
estimate the values of $\alpha$, $\beta$ and $\gamma$,
we will sketch it here. It is based on the following Lemma which
indicates  why the conditions $\mathfrak{A}$ need to be imposed.
\begin{lem}\label{lem:1}
  Let the function
  $h\left(\vec{\vphi}\right):\mathbb{R}^{n}\rightarrow\mathbb{R}$
  satisfy the following conditions:
  $h\in C\left(\mathbb{R}^{n}\right)$ with
  $C\left(\mathbb{R}^{n}\right)$ denoting the space of continuous 
  functions over $\mathbb{R}^{n}$, 
  $h$ is twice differentiable at $\vec{0}$ with $h\left(\vec{0}\right)=0$
  and $\left.\nabla_{\vec{\vphi}}h\left(\vec{\vphi}\right)\right|_{\vec{\vphi}=\vec{0}}=\vec{0}$
  and $\exists \,K,L<\infty:\|\vec{\vphi}\|\geq L\Rightarrow
  h\left(\vec{\vphi}\right)\leq K\|\vec{\vphi}\|^{2}$.
  Then:
  \[
  \sup_{\vec{\vphi}\in\mathbb{R}^{n}}
  \frac{h\left(\vec{\vphi}\right)}{\|\vec{\vphi}\|^{2}} < \infty
  \]
\end{lem}
When using Lemma~\ref{lem:1}, two problems may arise in ensuring that the
supremum is finite.
(i) Small values of $\|\vec{\vphi}\|$ create a small denominator which may
lead to large values of $\frac{h\left(\vec{\vphi}\right)}{\|\vec{\vphi}\|^{2}}$
near $\vec{\vphi}=\vec{0}$. This is eliminated by the conditions of
Lemma~\ref{lem:1}, $h\left(\vec{0}\right)=0$  and
$\left.\nabla_{\vec{\vphi}}h\left(\vec{\vphi}\right)\right|_{\vec{\vphi}=\vec{0}}=\vec{0}$.
We show in Appendices~\ref{app:A} and \ref{app:BC} that all quantities
taking the role of $h(\vec{\vphi})$ satisfy these conditions so that
we can employ Lemma~\ref{lem:1}.
(ii) Large values of $\|\vec{\vphi}\|$ may cause large values of
$h\left(\vec{\vphi}\right)$ which in turn can lead to  large
values of $\frac{h\left(\vec{\vphi}\right)}{\|\vec{\vphi}\|^{2}}$ as
$\|\vec{\vphi}\|\rightarrow\infty$.
The second problem is handled by imposing that
$h\left(\vec{\vphi}\right)$ grows at most quadratically with
$\|\vec{\vphi}\|$ which is
guaranteed by the conditions $\mathfrak{A}$.
Continuity of $h(\vec{\varphi})$ then ensures finiteness of the supremum's
argument for all intermediate values of $\|\vec{\vphi}\|$, $0 <
\|\vec{\vphi}\| < \infty$.  

\subsection{Final-time contribution to the second order ansatz --
proof  of  existence of finite $A$}
\label{app:A}
Without any assumption on the normalization, i.e. without any
reference to quantum control, 
we verify that $A$ of Eq.~(\ref{eq:defA})
is well-defined with $A<\infty$ based on the conditions $\mathfrak{A}$
and  using Lemma~\ref{lem:1}.
Let $\vec{\psi}$ denote the change in the state vector.
Specifically, we have to check that, for
\[
h\left(\vec{\psi}\right)=\vec{\psi}\cdot\Delta\vec{f}
\]
with $\Delta f$ defined in Eq.~(\ref{eq:deltaf}),
(i) $h$ is twice differentiable at $\vec{0}$ with
$h\left(\vec{0}\right)=0$, 
(ii)
$\left.\nabla_{\vec{\psi}}h\left(\vec{\psi}\right)\right|_{\vec{\psi}=\vec{0}}=\vec{0}$
and (iii) $h\left(\vec{\psi}\right)$ is bounded, i.e.
finite constants $K$ and $L$ exist such that
$\exists \; K,L < \infty : \|\vec{\psi}\|\geq L\Rightarrow h\left(\vec{\psi}\right)\leq
K\|\vec{\psi}\|^{2}$. Obviously $h$ is twice differentiable since $\vec{f}$
is twice differentiable and $\vec{\chi}(T)$ is a constant with respect to
$\vec{\psi}$. Since
$\left.J_T\left(\vec{\vphi}^{(i)}+\vec{\psi}\right)\right|_{\vec{\psi}=0}
-J_T\left(\vec{\vphi}^{(i)}\right) =
J_T\left(\vec{\vphi}^{(i)}\right)-J_T\left(\vec{\vphi}^{(i)}\right)=0$,
we find $h\left(\vec{0}\right)=0$.
We check whether
$\left.\nabla_{\vec{\psi}}h\left(\vec{\psi}\right)\right|_{\vec{\psi}=\vec{0}}=\vec{0}$,
\begin{eqnarray*}
\left.\nabla_{\vec{\psi}}h\left(\vec{\psi}\right)\right|_{\vec{\psi}=\vec{0}}
&=& \left(\frac{\vec{\psi}}{\|\vec{\psi}\|}\right)\cdot\vec{\chi} \\
&&+
\left.\frac{\partial  J_T\left(\vec{\vphi}^{(i)}+\vec{\psi}\right)}
  {\partial\vec{\psi}}
\right|_{\vec{\psi}=\vec{0}}\cdot\left(\frac{\vec{\psi}}{\|\vec{\psi}\|}\right)\\
&=&\vec{\chi}\cdot\hat{e}_{\vec{\psi}}+\frac{\partial
  J_T}{\partial\vec{\vphi}}\left(\vec{\vphi}^{(i)}\right)\cdot\hat{e}_{\vec{\psi}}  \,.
\end{eqnarray*}
Since we have to check these conditions at $t=T$ and remembering that
$\vec{\chi}\left(T\right)=-\frac{\partial
  J_T}{\partial\vec{\vphi}}\left(\vec{\vphi}^{(i)}\left(T\right)\right)$, we
obtain the desired result,
\begin{eqnarray*}
\left.\nabla_{\vec{\psi}}h\left(\vec{\psi}\right)\right|_{\vec{\psi}=\vec{0},t=T}
&=& \vec{\chi}\cdot\hat{e}_{\vec{\psi}\left(T\right)}+\frac{\partial
  J_T}{\partial\vec{\vphi}}
\left(\vec{\vphi}^{(i)}\left(T\right)\right)\cdot\hat{e}_{\vec{\psi}\left(T\right)}\\
&=&\vec{\chi}\cdot\hat{e}_{\vec{\psi}\left(T\right)}-
\vec{\chi}\cdot\hat{e}_{\vec{\psi}\left(T\right)}\\
&=&\vec{0}  \,.
\end{eqnarray*}
Finally, the condition set $\mathfrak{A}$ tells us that
$J_T\left(\vec{\vphi}\right)
\overset{\|\vec{\psi}\|\rightarrow\infty}{=}O\left(\|\vec{\psi}\|^{2}\right)$,
hence
\begin{eqnarray*}
h\left(\vec{\psi}\right) &=&
\left(\vec{\chi}\cdot\vec{\psi}\right)+J_T\left(\vec{\vphi}^{(i)}+\vec{\psi}\right)-J_T\left(\vec{\vphi}^{(i)}\right)\\
&\overset{\|\vec{\psi}\|\rightarrow\infty}{=}&
O\left(\|\vec{\psi}\|\right)+O\left(\|\vec{\psi}\|^{2}\right)+
O\left(1\right)\\ &=&O\left(\|\vec{\psi}\|^{2}\right)
\end{eqnarray*}
and condition (iii)
is fulfilled. We may thus use Lemma~\ref{lem:1} which guarantees the
existence of $A$.

\subsection{Intermediate-time contribution to the second order ansatz
  -- proof of  existence of  finite $B$, $C$}
\label{app:BC}

Analogously to the previous section,
we now verify that $B$ and $C$ are well-defined with $0<B<\infty$
and $C>-\infty$
without any assumption on the normalization of the state vector.
The constant $B$ defined in Eq.~(\ref{eq:defB})
is easily checked using Lemma~\ref{lem:1}.
Similarly to proving the existence of finite $A$,
we have to check whether the conditions for  Lemma~\ref{lem:1} are
fulfilled
for
\[
h\left(\vec{\psi}(t)\right)=\vec{\psi}(t)\cdot\Delta\vec{f}
\]
In analogy to Appendix~\ref{app:A}, it is obvious that  $h$
is twice differentiable at $\vec{0}$ with $h\left(\vec{0}\right)=0$.
In order to check that
$\left.\nabla_{\vec{\psi}(t)}h\left(\vec{\psi}(t)\right)\right|_{\vec{\psi}(t)=\vec{0}}=0$,
we use the product rule,
\begin{widetext}
\begin{eqnarray*}
\left.\nabla_{\vec{\psi}(t)}h\left(\vec{\psi}(t)\right)\right|_{\vec{\psi}(t)=\vec{0}}
=
\left(\frac{\vec{\psi}(t)}{\|\vec{\psi}(t)\|}\right)\cdot
\underbrace{\left.\Delta\vec{f}\left(\vec{\psi}(t)\right)\right|_{\vec{\psi}(t)=0}}_{=\vec{0}}
+\underbrace{\left.\vec{\psi}(t)\cdot
    \nabla_{\vec{\psi}(t)} \left(\Delta\vec{f}\right)\right|_{\vec{\psi}(t)=0}}_{=\vec{0}}
=\vec{0}  \,.
\end{eqnarray*}
\end{widetext}
Finally, the condition set $\mathfrak{A}$ tells us that
$\|\vec{f}\left(\vec{\vphi}\right)\|
\overset{\|\vec{\psi}\|\rightarrow\infty}{=}O\left(\|\vec{\psi}\|\right)$,
hence
\[
h\left(\vec{\psi}\right)=
\vec{\psi}\cdot\Delta\vec{f}
\overset{\|\vec{\psi}\|\rightarrow\infty}{=}O\left(\|\vec{\psi}\|^{2}\right)\,,
\]
i.e. $h$ is indeed quadratically bounded.
We may thus use Lemma~\ref{lem:1} which guarantees the existence of
$B < \infty$. Moreover,
$B>0$ due to taking the square modulus inside the supremum.

To prove the existence of finite $C$ defined by Eq.~(\ref{eq:defC}),
we distinguish the two cases for large and small $\|\vec{\psi}\|$.
For large $\|\vec{\psi}(t)\|$, the argument of the infimum is
finite. This follows from  $\dot{\vec{\chi}}(t)$ and $\vec{\chi}(t)$ being
constant with respect to $\vec{\psi}(t)$ and the fact that
for large $\|\vec{\vphi}\|$, we have $\|\vec{f}\|\leq K_{1}\|\vec{\vphi}\|$
and $\left|g\right|\leq K_{2}\|\vec{\vphi}\|^{2}$.
Hence
\[
\dot{\vec{\chi}}(t)\cdot\vec{\psi}(t)+\vec{\chi}(t)\cdot\Delta\vec{f}-\Delta
g\overset{\|\vec{\psi}(t)\|
  \rightarrow\infty}{\leq}O\left(\|\vec{\psi}(t)\|^{2}\right)\,.
\]
For small $\|\vec{\psi}(t)\|$ we have to check that the zeroth and first
order terms in the denominator disappear.
Inserting the expression
for $\dot{\vec{\chi}}$ into the definition of $C$ yields
\begin{widetext}
  \begin{equation}
    \label{eq:infC}
    C=\inf_{\vec{\psi}(t)\in\mathbb{R}^{2NM};t\in\left[0,T\right]}
    \frac{\left(-\nabla_{\vec{\vphi}(t)}\vec{f}^{\,T} \cdot\vec{\chi}(t)\right)
      \cdot\vec{\psi}(t)
      +\vec{\chi}(t)\cdot\Delta\vec{f} +
      \nabla_{\vec{\vphi}} g \cdot\vec{\psi}(t)-\Delta  g}
    {\left(\vec{\psi}(t)\cdot\vec{\psi}(t)\right)}\,,
  \end{equation}
\end{widetext}
For sufficiently small $\varepsilon$ and $\|\vec{\psi}(t)\|<\varepsilon$, we
may approximate $\Delta f$ and $\Delta g$  to the first order,
\begin{eqnarray*}
\Delta\vec{f} &\simeq&
\nabla_{\vec{\vphi}(t)}\vec{f}\cdot\vec{\psi}(t)
+O\left(\|\vec{\psi}\|^{2}\right) \,,\\
\Delta g &\simeq&
\nabla_{\vec{\vphi}(t)} g\cdot\vec{\psi}(t)
+O\left(\|\vec{\psi}\|^{2}\right)  \,.
\end{eqnarray*}
This yields
\begin{widetext}
\begin{eqnarray*}
&&\frac{\left(-\nabla_{\vec{\vphi}(t)}\vec{f}^{\,T}\cdot\vec{\chi}(t)\right)
  \cdot\vec{\psi}(t)+\nabla_{\vec{\vphi}(t)} g \cdot\vec{\psi(t)}
  +\vec{\chi}(t)\cdot\Delta\vec{f}-\Delta g}{\left(\vec{\psi}(t)\cdot\vec{\psi}(t)\right)}\\
&&\quad\quad\overset{\|\vec{\psi}(t)\|\rightarrow0}{\longrightarrow}
\frac{\left(-\nabla_{\vec{\vphi}(t)}\vec{f}^{\,T}
    \cdot\vec{\chi}(t)\right)\cdot\vec{\psi}(t)
  +\nabla_{\vec{\vphi}(t)} g \cdot\vec{\psi}(t)
  +\vec{\chi}(t)\cdot\left(\nabla_{\vec{\vphi}(t)}\vec{f} \cdot\vec{\psi}(t)\right)
  -\nabla_{\vec{\vphi}(t)} g  \cdot\vec{\psi}(t)
  +O\left(\|\vec{\psi}(t)\|^{2}\right)}{\left(\vec{\psi}(t)\cdot\vec{\psi}(t)\right)}\\
& &\quad\quad \overset{\|\vec{\psi}(t)\|\rightarrow0}{\longrightarrow}
 \frac{O\left(\|\vec{\psi}(t)\|^{2}\right)}
 {\left(\vec{\psi}(t)\cdot\vec{\psi}(t)\right)}\\ &&\quad\quad=O\left(1\right)\,,
\end{eqnarray*}
\end{widetext}
i.e. $C$ remains finite also for small $\|\vec{\psi}(t)\|$.

\section{Applying Krotov's proof to quantum control problems}
\label{sec:adaptproof}

We adapt Konnov and Krotov's results~\cite{Konnov99}
to the case that $\vec{\vphi}$ is composed of 
the $M$ real and imaginary expansion coefficients of $N$ normalized
quantum state vectors.
In quantum control applications, 
some of the conditions $\mathfrak{A}$, cf.
Appendix~\ref{sec:proof},  are trivially fulfilled.
Specifically, the state vector $\vec{\vphi}$ is
inherently bounded for any field $\epsilon$ if $\vec{\vphi}$ is
obtained from the equation of motion for a given $\epsilon$
due to normalization.
In particular, $\|\vec{\vphi}\|$ cannot become larger than
$\sqrt{N}$ and  we may reduce the candidate space
$\mathbb{R}^{2NM}$ for $\vec{\vphi}$ to a \textit{compact} subset
 of $\mathbb{R}^{2NM}$, namely $\mathbb{X}=\left\{
  \left.\vec{\vphi}\in\mathbb{R}^{2NM}\right|\|\vec{\vphi}\|\leq \sqrt{N}\right\}
\subset\mathbb{R}^{2NM}$.
The conditions concerning the behavior
for large $\|\vec{\vphi}\|$ are then trivially fulfilled
because the complete space of interest
$[0,T]\times\mathbb{X}\times\mathbb{E}$, that is the set of all
$\{\left(t,\vec{\vphi},\epsilon\right)\}$,
is compact. 
The boundedness of the right-hand side of the equations
of motion, the Jacobian and the functionals
is then guaranteed by simply asking
$f\left(t,\vec{\vphi},\epsilon\right)$,  $\boldsymbol{\mathsf{J}}$,
$J_T(\vec{\vphi})$ and
$g\left(\epsilon,\vec{\vphi},t\right)$ to be \textit{regular}.

\begin{figure}[tb]
  \centering
  \includegraphics[width=0.7\linewidth]{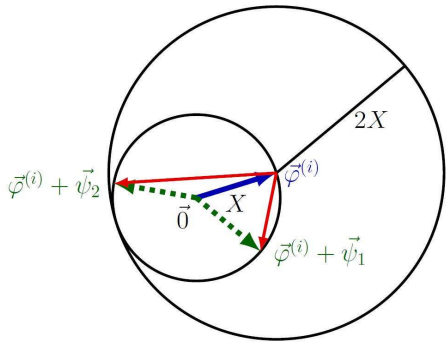}
  \caption{All admissible states, $\vec{\vphi}^{(i)}$ and
    $\vec{\vphi}^{(i+1)}=\vec{\vphi}^{(i)}+\vec{\psi}$
    lie on a sphere of radius $X=\sqrt{N}$ around the
    origin. Therefore the
    norm of the change in the states states, $\|\vec{\psi}\|$, varies
    between 0 and $2X$.
  }
  \label{fig:norm}
\end{figure}

Due to the restriction of the states to the compact subset
$\mathbb{X}\subset\mathbb{R}^{2NM}$,
the proof simplifies for quantum control
applications which we use to obtain
straightforward estimates of the constants $\alpha$, $\beta$
and $\gamma$.
The changes in $R$ and $G$ due to variation of the state 
from the extremal point, $\vec{\vphi}^{(i)}$, to all possible states,
$\vec{\vphi}^{(i)}+\vec{\psi}$, is measured by
\begin{eqnarray}
\Delta G\left(\vec{\psi}\right) & = &
G\left(\vec{\vphi}^{(i)}(T)+\vec{\psi}\right)
-G\left(\vec{\vphi}^{(i)}(T)\right) \,, \label{eq:DeltaG}\\
\Delta R\left(\vec{\psi}(t),t\right) & = &
R\left(\vec{\vphi}^{(i)}(t)+\vec{\psi}(t),\epsilon^{(i)}(t),t\right)\nonumber\\
&&-R\left(\vec{\vphi}^{(i)}(t),\epsilon^{(i)}(t),t\right)\,. \label{eq:DeltaR}
\end{eqnarray}
Then the global extremum conditions,
Eqs.~(\ref{eq:minR})-(\ref{eq:maxG}), 
correspond to $\Delta G\left(\vec{\psi}\right) \leq 0$ and
$\Delta R\left(\vec{\psi}(t),t\right)\geq 0$.
In quantum control applications, any state
$\vec{\vphi}^{(i)}+\vec{\psi}$ that is candidate for
$\vec{\vphi}^{(i+1)}$ must also be normalized. Geometrically, all
states $\vec{\vphi}^{(i)}$ and $\vec{\vphi}^{(i)}+\vec{\psi}$ lie
therefore on a 
sphere of radius $X$. The norm of the vectors $\vec{\psi}$ varies
between zero and $2X$ since 
for any two vectors $\vec{\vphi}^{(i)}$, $\vec{\vphi}^{(i+1)}$ in $\mathbb{X}$
the minimum distance is zero while the maximum distance is $2X$.
This is illustrated in Fig.~\ref{fig:norm}.
The space for the state change vectors $\vec{\psi}$ is then given by
$\mathbb{Y}=\left\{ \left.\vec{\psi}\in\mathbb{R}^{2NM}\right|
  \exists \vec{\vphi}^{(i)} \in \mathbb{X}:
  \vec{\vphi}^{(i)} +\vec{\psi} \in \mathbb{X}\right\} \subset\mathbb{R}^{2NM}$.
In quantum control applications it is thus sufficient
to vary the vector $\vec{\vphi}^{(i+1)}$ over the sphere of radius
$X$ around the origin instead of the full $\mathbb{R}^{2NM}$.

With the definitions of $A$, $B$ and $C$,
cf. Eqs.~\eqref{eq:defA}, \eqref{eq:defB},  \eqref{eq:defC},
a strict maximum condition for $G$  and minimum condition for $R$ is
transformed into
\begin{eqnarray}
  \label{eq:condA}
  \sigma(T) &<& -2A\,, \\
  \label{eq:condBC}
  \frac{1}{2}\dot{\sigma}\left(t\right)-
  \left|\sigma\left(t\right)\right|\cdot B+C&>&0\,.
\end{eqnarray}
A solution $\sigma(t)$ fulfilling these inequalities exists and it 
is straightforward to check that the Konnov and Krotov's ansatz
Eq.~\eqref{eq:sigma}
satisfies (\ref{eq:condA}) and (\ref{eq:condBC}). 
More generally, the ansatz~\cite{Konnov99}
\[
\sigma\left(t\right)=\alpha\left(e^{\gamma\left(T-t\right)}-1\right)+\beta
\]
with $\alpha,\beta<0$ and $\gamma>0$ leads to the inequalities
\begin{subequations}\label{eq:Condition set}
\begin{eqnarray}
\beta+2A & < & 0\,,  \\
-\frac{\alpha\gamma}{2}-B\beta+C & > & 0\,,
\end{eqnarray}
\end{subequations}
which have at least one specific solution, namely
Eq.~(\ref{eq:sigma}), with
\begin{eqnarray*}
\alpha  =  \frac{\bar{C}}{\bar{B}}-\bar{A} \,\,,\,\,
\beta  =  -\bar{A}\,\,,\,\,
\gamma  =  \bar{B}\,.
\end{eqnarray*}
Note that if a set $\left\{ \alpha_{0},\beta_{0},\gamma_{0}\right\} $
fulfills the inequalities~(\ref{eq:Condition set}), then
any other set $\left\{ \alpha,\beta,\gamma\right\} $
with $\alpha\leq\alpha_{0}<0$, $\beta\leq\beta_{0}<0$, $\gamma\geq\gamma_{0}>0$
does so, too. This flexibility can be utilized to estimate the
constants $A$, $B$ and $C$. Therefore 
 the suprema in Eqs.~(\ref{eq:defA}) and
(\ref{eq:defB}) and the infimum in Eq.~(\ref{eq:defC}) can be
 estimated analytically.

\section{Analytical estimate of the parameters of the second
  order contribution} 
\label{sec:ABC}

The arguments of the suprema, respectively of the
infimum, in $A$, $B$ and $C$ can be expressed in terms of Taylor
series starting at the first, respectively second, order. Evaluating
the remainder term of
these Taylor series, we obtain estimates for $A$, $B$ and $C$.
Let $W\left(\vec{\vphi}\right)$ be a
scalar, vector or matrix depending on $\vec{\vphi}$. $W_{n}\left(\vec{\vphi}\right)$
denotes the Taylor expansion of $W$ around $\vec{\vphi}^{(i)}$ starting
at the $n$-th order or, in other words, $W_{n}\left(\vec{\vphi}\right)$
equals $W\left(\vec{\vphi}\right)$ minus the first $n-1$ terms of its
Taylor expansion around $\vec{\vphi}^{(i)}$. For example, we obtain for
a scalar field
\begin{eqnarray*}
  W_{0}\left(\vec{\vphi}\right) & = & W\left(\vec{\vphi}^{(i)}\right)\,,\\
  W_{1}\left(\vec{\vphi}\right) & = &
  W\left(\vec{\vphi}\right)-W\left(\vec{\vphi}^{(i)}\right)\,,\\
  W_{2}\left(\vec{\vphi}\right) & = &
  W\left(\vec{\vphi}\right)   -W\left(\vec{\vphi}^{(i)}\right)\\
  &&-\nabla_{\vphi} W \left(\vec{\vphi}^{(i)}\right)\cdot
  \left(\vec{\vphi}-\vec{\vphi}^{(i)}\right)\,,
\end{eqnarray*}
and so forth.
The Taylor series starting at the $n$-th order can be approximated by
evaluating the remainder term. For a scalar field $W\left(\vec{\vphi}\right)$ this 
is given by
\begin{eqnarray}
  \label{eq:remainder}
  W_{n}\left(\vec{\vphi}\right) =
  \mathcal{R}^W_{\vec{\vphi}^{(i)},n}\left(\vec{\psi}\right) =
  \sum_{\left|\alpha\right|=n}
  \frac{1}{\alpha!} \partial^{\alpha} W
    \left(\vec{\vphi}^{(i)}+c\vec{\psi}\right)
    \vec{\psi}^{\alpha}
\end{eqnarray}
for a $c\in\left(0,1\right)$ and with 
$\vec{\psi} = \vec{\vphi}- \vec{\vphi}^{(i)}$.
Here $\alpha$ is a multi-index representing the
$2NM$-tuple of natural numbers including zero, 
\begin{equation}
\label{eq:alpha}
  \left|\alpha\right|  =  \sum_{i=1}^{2NM}\alpha_{i}\,,\,
  \frac{1}{\alpha!} =  \frac{1}{\prod_{i=1}^{2NM}\alpha_{i}!}\,,\,
  \vec{\psi}^{\alpha}  =  \prod_{i=1}^{2NM}\psi_i^{\alpha_{i}}\,,
\end{equation}
and
\begin{equation}
  \label{eq:dalpha}
  (\partial^{\alpha} W)\left(\vec{\vphi}^{(i)}\right)  =
  \prod_{i=1}^{2NM}\left.\frac{\partial^{\alpha_{i}} W\left(\vec{\vphi}\right)}{\partial
      \vphi_{i}^{\alpha_{i}}}\right|_{\vec{\vphi}=\vec{\vphi}^{(i)}}\,.  
\end{equation}
The remainder can be estimated by
\begin{equation}
  \label{eq:estimateremainder}
  \mathcal{R}^W_{\vec{\vphi}^{(i)},n}\left(\vec{\psi}\right) \le
  \frac{1}{\alpha!}
  M^W_n\left(\vec{\vphi}^{(i)}\right) \vec{\psi}^\alpha \;,|\alpha|=n\,,
\end{equation}
with
\[
M^W_n \left(\vec{\vphi}^{(i)}\right) =
\sup_{\vec{\psi}\in\mathbb{Y}; \left|\alpha\right|=n}
\partial^{\alpha}
  W\left(\vec{\vphi}^{(i)}+\vec{\psi}\right)\,.
\]
An estimate that is independent of the state
$\vec{\vphi}^{(i)}$ is obtained by taking the supremum over all
possible $\vec{\vphi}^{(i)}$, i.e. we define
\begin{eqnarray}
  \label{eq:defM}
  M^W_n &\equiv &
  \sup_{\vec{\vphi}^{(i)}\in\mathbb{X}}M_n\left(\vec{\vphi}^{(i)}\right)
  \nonumber \\
  &=& \nonumber \sup_{\vec{\vphi}^{(i)}\in\mathbb{X};
      \vec{\psi}\in\mathbb{Y}; \left|\alpha\right|=n}
  \partial^{\alpha} W\left(\vec{\vphi}^{(i)}+\vec{\psi}\right) \\
  &=&\sup_{\vec{\Psi}\in\mathbb{X};\left|\alpha\right|=n}
  \partial^{\alpha} W\left(\vec{\Psi}\right)\,.
\end{eqnarray}
This method to estimate the Lagrange remainder term lends itself to an
intuitive geometrical interpretation, cf. Fig.~\ref{fig:norm}.
For given $\vec{\vphi}^{(i)}$, the Taylor expansion of $W$ around
$\vec{\vphi}^{(i)}$ starting at the $n$th order, $W_{n}$,
can be estimated by the supremum of the $n$th derivatives taken over
the sphere around this state with radius $\| \vec{\psi}\| =y_{0}$.
To give an estimate of $W_{n}$
that holds for any $\vec{\vphi}^{(i)}$, we have to calculate the supremum
around all possible state vectors. Since all $\vec{\vphi}^{(i)}$
are located on a sphere around the origin with radius $\sqrt{N}$, we simply
need to take  the supremum of the $n$th derivatives
over all vectors within a ball around the origin with radius
$2\sqrt{N}$. Since the Lagrange form of the remainder is based on the
mean value theorem, the difference between two values of a function
$W\left(\vec{\vphi}^{(i)}+\vec{\psi}\right)-W\left(\vec{\vphi}^{(i)}\right)$
is estimated by the first derivative of the function at some point
between $\vec{\vphi}^{(i)}+\vec{\psi}$ and $\vec{\vphi}^{(i)}$. Since
both $\vec{\vphi}^{(i)}$  and  $\vec{\vphi}^{(i)}+\vec{\psi}$ are
located on the sphere of radius $\sqrt{N}$ around the origin, any
difference of a function between these two points can only
concern values of derivatives \textit{inside} this sphere. Therefore
we can restrict the supremum to be taken over all states
in $\mathbb{X}$ in the last line of Eq.~\eqref{eq:defM}. 

We now apply the estimate of the Lagrange remainder to derive the
constants $A$, $B$ and $C$ and use the braket notation in the following. 
Considering in Eq.~\eqref{eq:defA} the Taylor expansion of
$J_T\left(\{\vphi_k^{(i)}(T)+\Delta\vphi_k\}\right)$ in $\Delta\vphi_k$
around $\vphi_k^{(i)}(T)$, the first order term cancels with
$\langle\chi_k(T)|\Delta\varphi_k(T)\rangle + c.c.$ 
since $\Ket{\chi_k(T)}$ is given in
terms of the gradient of $J_T$, cf. Eq.~\eqref{eq:chiT}.
The zeroth order term is nullified by 
$-J_T\left(\{\vphi_k^{(i)}(T)\right)$,
\begin{widetext}
\begin{eqnarray*}
A & = &
\sup_{\{\Delta\vphi_k\}}
\frac{J_{T}(\{\vphi_k^{(i)}(T)+\Delta\vphi_k(T)\})
- J_T(\{\vphi_k^{(i)}(T)\}) + \sum_{k=1}^N \left[ \langle\chi_k(T)|\Delta\varphi_k(T)\rangle
+ \langle\Delta\varphi_k(T)|\chi_k(T)\rangle \right]}
{\sum_{k=1}^{N} \Braket{\Delta\vphi_k(T)|\Delta\vphi_k(T)}} \\
& = & \sup_{\{\Delta\vphi_k\}}
\frac{J_{T,2}(\{\Delta\vphi_k(T)\})}
{\sum_{k=1}^{N}\Braket{\Delta\vphi_k(T)|\Delta\vphi_k(T)}} \,.
\end{eqnarray*}
\end{widetext} 
The argument of the supremum in the 
definition of $A$, Eq.~\eqref{eq:defA},
can therefore be viewed as the Taylor expansion of
$J_T\left(\{\vphi_k^{(i)}(T)+\Delta\vphi_k\}\right)$  around the 
$\vphi_k^{(i)}(T)$ starting at the second order, divided by
$\sum_{k=1}^{N}\Braket{\Delta\vphi_k(T)|\Delta\vphi_k(T)}$.
It is now possible to estimate $J_{T,2}$ by its Lagrange remainder 
according to Eq.~\eqref{eq:estimateremainder},
\begin{eqnarray*}
A & = & \sup_{\{\Delta\vphi_k\}}
\frac{\mathcal{R}^{J_T}_{\{\vphi_k^{(i)}(T)\},2}(\{\Delta\vphi_k(T)\})}
{\sum_{k=1}^{N}\Braket{\Delta\vphi_k(T)|\Delta\vphi_k(T)}} \\
& \leq & \sup_{\{\Delta\vphi_k\}}
\frac{\frac{1}{2}
  M^{J_T}_2\left(\{\vphi_k^{(i)}(T)\}\right) 
\sum_{k=1}^{N}\Braket{\Delta\vphi_k(T)|\Delta\vphi_k(T)}}
{\sum_{k=1}^{N}\Braket{\Delta\vphi_k(T)|\Delta\vphi_k(T)}}\\
&=& \frac{1}{2}
  M^{J_T}_2\left(\{\vphi_k^{(i)}(T)\}\right)\\
&=& \frac{1}{2} \sup_{\{\Delta\vphi_k\};\left|\alpha\right|=2 }
  \partial^{\alpha} J_T({\{\Delta\vphi_k(T)}\})
  \,,
\end{eqnarray*}
yielding Eq.~\eqref{eq:A_est}.
Note that for functionals $J_T$ that are quadratic in the states, 
the \textit{global} convergence condition (\ref{eq:condA})
coincides with the local one,  $\nabla_{\vphi}^2 G < 0$, that was used
in Ref.~\cite{SklarzPRA02}. This can be
seen by inserting the ansatz for $\Phi$, Eq.~(\ref{eq:Phinew}), into
$\nabla_{\vphi}^2 G < 0$,
\[
\left\Vert \nabla_{\vphi}^2 J_T
  \left(\vphi^{(i)}(T)\right)\right
\Vert + \sigma(T) < 0 \,.
\]

To find an expression for the constant $B$ defined by
Eq.~\eqref{eq:defB}, we rewrite the change $\Delta f_k$
in the equations of motion due to changes in the states,
cf. Eq.~(\ref{eq:deltaf}), in terms of the Taylor expansion of the
Hamiltonian in $\Delta\vphi_k(t)$ around the
$\vphi_k^{(i)}(t)$  starting at the first order, 
$\Op H_1$, and obtain 
\begin{widetext}
\begin{eqnarray*}
  B &\leq& \sup_{\{\Delta\vphi_k\};t\in\left[0,T\right]}
  \left|\frac{\sum_{k=1}^{N}\left[
      \Braket{\Delta\vphi_k(t)|
        \Op H_{1}(\vphi_k+\Delta\vphi_k,\epsilon^{(i)})\
        |\vphi_k^{(i)}(t)} + 
      \Braket{\vphi_k(t)|
        \Op H^{\dagger}_{1}(\vphi_k+\Delta\vphi_k,\epsilon^{(i)})\
        |\Delta\vphi_k^{(i)}(t)} \right]}
    {\sum_{k=1}^{N}\Braket{\Delta\vphi_k(t)|\Delta\vphi_k(t)}}\right| \\
  && + 2 \sup_{\{\Delta\vphi_k\};t\in\left[0,T\right]}
  \left|\frac{\sum_{k=1}^N \mathfrak{Im} 
    \Braket{\Delta\vphi_k(t)|\Op H\left(\vphi_k^{(i)}(t)+\Delta\vphi_k(t),\epsilon^{(i)},t\right)|\Delta\vphi_k(t)}}
    {\sum_{k=1}^{N}\left[\Braket{\Delta\vphi_k(t)|\Delta\vphi_k(t)}\right]}\right|.
\end{eqnarray*}
Using the Cauchy Schwarz inequality for the scalar products
in the argument of the first supremum and Eq.~\eqref{eq:defM} for the
second supremum yields
\begin{eqnarray*}
  B &\leq &
   \sup_{\{\Delta\vphi_k\};t\in\left[0,T\right]}
   \left(\frac{2}{\sum_{k=1}^{N}\Braket{\Delta\vphi_k(t)|\Delta\vphi_k(t)}}
     \sum_{k=1}^{N}\left[\sqrt{\Braket{\Delta\vphi_k(t)|\Delta\vphi_k(t)}} \cdot\left\|
         \Op H_{1}\left(\vphi_k+\Delta\vphi_k,\epsilon^{(i)}\right)\right\|\cdot\sqrt{
         \langle\vphi_k^{(i)}(t)|\vphi_k^{(i)}(t)\rangle}
     \right]\right) \\
  &&+ 2 \sup_{\{\Delta\vphi_k\};t\in\left[0,T\right]}
  \left|\frac{\sum_{k=1}^N \mathfrak{Im} 
    \Braket{\Delta\vphi_k(t)|\Op H\left(\Delta\vphi_k(t),\epsilon^{(i)},t\right)|\Delta\vphi_k(t)}}
    {\sum_{k=1}^{N}\Braket{\Delta\vphi_k(t)|\Delta\vphi_k(t)}}\right|\,.
\end{eqnarray*}
\end{widetext}
To evaluate the first supremum, we estimate the Taylor expansion of
the Hamiltonian starting at the first order, $\Op H_1$, by its
Lagrange remainder, 
\begin{eqnarray}
\left\|\Op H_1\left(\Delta\vphi_k\right)\right\| &\le & 
\sum_{k=1}^{N}M^{|\Op H|,k}_1\cdot
\sqrt{\Braket{\Delta\vphi_k|\Delta\vphi_k}} \nonumber \\
&=&  \sum_{k=1}^{N}\sup_{\Delta\vphi_k;|\alpha|=1}\left|
\partial^{\alpha} \Op H\left(\Delta\vphi_k\right)\right|
\nonumber \\
&&\quad\quad\cdot
\sqrt{\Braket{\Delta\vphi_k|\Delta\vphi_k}}\,.\label{eq:w1},
\end{eqnarray}
Note that the  absolute value of the derivatives is required in the
estimation of the Lagrange remainder since we need to estimate the
norm of $\Op H_1$.
With 
$\sum_{k=1}^{N}\sqrt{\langle\vphi_k^{(i)}(t)|\vphi_k^{(i)}(t)\rangle}=\sqrt{N}$,
we obtain Eq.~\eqref{eq:Bfinal} for $B$. 
In particular, for 
Hamiltonians that do not depend on the state, 
$\Op H\left(\vphi_k^{(i)}(t)+\Delta\vphi_k(t),\epsilon^{(i)}\right)
  -\Op H \left(\vphi_k^{(i)}(t),\epsilon^{(i)}\right) = 0$, and
the first supremum can taken to be zero. Then $B$ is given by the
second supremum alone which is twice the maximum absolute value of
the imaginary part the Hamiltonian's eigenvalues. That is, the second
supremum is non-zero only for non-unitary time
evolution. In summary, for linear equations of motion and unitary time
evolution, $B=0$. The differential inequality system for
$\sigma(t)$ then reduces to 
\begin{eqnarray*}
\frac{1}{2}\sigma(T)+A&<&0 \,,\\
\frac{1}{2}\dot{\sigma}(t)+C&>&0 \,,
\end{eqnarray*}
such that we obtain a
linear solution for $\sigma(t)$,
\begin{equation}
\sigma(t)=\bar{C}(T-t)-\bar{A}\,.
\end{equation}

To estimate the constant $C$ defined by Eq.~\eqref{eq:defC}, we
rewrite it, using Eq.~\eqref{eq:infC}, 
\begin{widetext}
  \begin{equation}
    \label{eq:Cini}
C = \inf_{\{\Delta\vphi_k\};t\in\left[0,T\right]}
\frac{\sum_{k=1}^{N}\left[
    \Braket{\chi_k^{(i)}(t)|f_{k,2}\left(\vphi_k^{(i)}+\Delta\vphi_k,\epsilon^{(i)}\right)}
  +\Braket{f_{k,2}\left(
      \vphi_k^{(i)}+\Delta\vphi_k,\epsilon^{(i)}\right)|\chi_k^{(i)}(t)}\right]
  -g_{2}\left(\{\vphi_k^{(i)}+\Delta\vphi_k\},t\right)}
{\sum_{k=1}^{N}\Braket{\Delta\vphi_k(t)|\Delta\vphi_k(t)}}\,,
  \end{equation}
where $\Ket{f_{k,2}}$ and $g_{2}$ are the Taylor expansions of $\Ket{f_k}$
and $g$ in $\Delta\vphi_k(t)$ around $\vphi_k^{(i)}(t)$ starting
at the second order.
Expressing $\Ket{f_{k,2}}$ in terms of the Taylor expansion of the
Hamiltonian starting at first order, 
$\Ket{f_{k,2}} = \Op
H_1\left(\vphi_k^{(i)}+\Delta\vphi_k,\epsilon^{(i)}\right)
|\Delta\vphi_k(t)\rangle$, and introducing the approximation
\begin{eqnarray*}
  -C &\ge& \sup_{\{\Delta\vphi_k\};t\in\left[0,T\right]}
-\frac{\sum_{k=1}^{N}\left[\Braket{\chi_k^{(i)}(t)|
\Op H_1\left(\vphi_k^{(i)}+\Delta\vphi_k,\epsilon^{(i)}\right)|\Delta\vphi_k(t)}
+\Braket{\Delta\vphi_k(t)|
\Op H^+_1\left(\vphi_k^{(i)}+\Delta\vphi_k,\epsilon^{(i)}\right)
|\chi_k^{(i)}(t)}\right]}
{\sum_{k=1}^{N}\Braket{\Delta\vphi_k(t)|\Delta\vphi_k(t)}} \\ &&+
\sup_{\{\Delta\vphi_k\};t\in\left[0,T\right]}
\frac{g_2\left(\{\vphi_k^{(i)}+\Delta\vphi_k\},t\right)}
{\sum_{k=1}^{N}\Braket{\Delta\vphi_k(t)|\Delta\vphi_k(t)}}\,,
\end{eqnarray*}
\end{widetext}
we may reuse our results for $\Op H_1$, cf. Eq.~\eqref{eq:w1},
together with $M^{-\Op H,k}_1 = -M^{\Op H,k}_1$
to estimate the first term. The estimation of the second term
involving $g_2$ proceeds analogously
to that of $J_{T,2}$, with $M^g_2$ given by
\[
M^g_2 = \sup_{\begin{subarray}{c} \{\Delta\vphi_k\};t\in[0,T] \\ |\alpha|=2 \end{subarray}}
\partial^{\alpha}g\left(\{\Delta\vphi_k\},t\right) \,.
\]
We thus obtain Eq.~\eqref{eq:Cfinal} for $C$.
For Hamiltonians that depend on the state, the first term in the
right-hand side of Eq.~(\ref{eq:Cfinal}) is non-zero.  Note
that the norm of all adjoint vectors $\chi_k^{(i)}$ is equal to $\sqrt{N}$
only for state-independent constraint $g$. For state-dependent
constraint $g$, $\chi_k^{(i)}$ is the
solution of an inhomogeneous Schr\"odinger equation and its norm may
be smaller or larger than $\sqrt{N}$. 
However, the norm of all adjoint vectors $\chi_k^{(i)}$ is
always known since the $\chi_k^{(i)}$ are calculated in the
previous iteration step, $i$.
So in order to estimate the first term of the right-hand
side of  Eq.~(\ref{eq:Cfinal}), we only need to determine
$\left\Vert\partial \Op H\right\Vert$ which is also needed for 
estimating $B$. In this case, $-C$ is then
given by the sum of the spectral radius of the operator
$\left\Vert\partial \Op H\right\Vert$ and the eigenvalue of
$\Op{D}(t)$ with largest magnitude.


\end{document}